\documentclass[twocolumn]{aastex63}

\usepackage{graphicx,xcolor}
\usepackage{lipsum}
\usepackage{gensymb}
\usepackage{placeins}
\usepackage{float}
\usepackage{hyperref}
\usepackage{threeparttablex}
\usepackage{longtable}
\usepackage{booktabs,multirow,tabularx} 
\usepackage{amsmath}  
\usepackage{graphicx} 
\usepackage{tabulary,capt-of}
\usepackage{booktabs}
\usepackage{rotating}
\usepackage{booktabs} 
\usepackage{tablefootnote}
\usepackage{natbib}
\makeatletter
\newcommand\footnoteref[1]{\protected@xdef\@thefnmark{\ref{#1}}\@footnotemark}
\makeatother
\usepackage{float}
\usepackage{lineno}

\newcommand{\mc}[1]{{#1}}

\shorttitle{NEP-NuSTAR $N_{{\mathrm{H}}}$ Distribution}
\shortauthors{Creech et al.}

\begin{document}

\title{The $N_{{\mathrm{H}}}$ Distribution of Hard X-ray Selected AGN in the NEP Field}


\author[0000-0002-9041-7437]{Samantha Creech}
\affiliation{Department of Physics \& Astronomy, The University of Utah, 115 South 1400 East, Salt Lake City, UT 84112, USA}

\author[0000-0002-2115-1137]{Francesca Civano}
\affiliation{NASA Goddard Space Flight Center}

\author[0000-0001-9110-2245]{Daniel R. Wik}
\affiliation{Department of Physics \& Astronomy, The University of Utah, 115 South 1400 East, Salt Lake City, UT 84112, USA}

\author[0000-0001-6564-0517]{Ross Silver}
\affiliation{NASA Goddard Space Flight Center}

\author[0000-0002-7791-3671]{Xiurui Zhao}
\affiliation{Department of Astronomy, University of Illinois at Urbana-Champaign, Urbana, IL 61801, USA}
\affiliation{Cahill Center for Astrophysics, California Institute of Technology, 1216 East California Boulevard, Pasadena, CA 91125, USA}

\author[0000-0002-6150-833X]{Rafael Ortiz III}
\affiliation{School of Earth and Space Exploration, Arizona State University,
Tempe, AZ 85287-1404, USA}

\author[0000-0001-8211-3807]{Tonima Ananna}
\affiliation{Department of Physics and Astronomy, Wayne State University, Detroit, MI 48202, USA}

\author[0000-0001-9440-8872]{Norman A. Grogin}
\affiliation{Space Telescope Science Institute, 3700 San Martin Drive, Baltimore, MD 21218, USA}

\author[0000-0003-1268-5230]{Rolf Jansen}
\affiliation{School of Earth and Space Exploration, Arizona State University, Tempe, AZ 85287-1404, USA}

\author[0000-0002-6610-2048]{Anton M. Koekemoer}
\affiliation{Space Telescope Science Institute, 3700 San Martin Drive,
Baltimore, MD 21218, USA}

\author[0000-0001-9262-9997]{Christopher N.A. Willmer} 
\affiliation{Steward Observatory, University of Arizona, 
933 North Cherry Avenue, Tucson, AZ 85721, USA}

\author[0000-0001-8156-6281]{Rogier A. Windhorst}
\affiliation{School of Earth and Space Exploration, Arizona State University,
Tempe, AZ 85287-1404, USA}

\defcitealias{Zhao21_cycle5}{Z21}
\defcitealias{Zhao24-cycle5-6}{Z24}

\begin{abstract}

X-ray surveys are one of the most unbiased methods for detecting Compton Thick (CT; $N_{{\mathrm{H}}} \geq 10^{24}$~cm$^{-2}$) AGN, which are thought to comprise up to $60\%$ of AGN within $z \lesssim 1.0$.
These CT AGN are often difficult to detect with current instruments, but the X-ray data within the JWST-North Ecliptic Pole (NEP)
Time Domain Field (TDF) present a unique opportunity to study faint and obscured AGN.
The NEP contains the deepest NuSTAR survey to date, and \citet{Zhao24-cycle5-6} detected 60 hard X-ray sources from the combined exposure of NuSTAR's Cycle 5 and 6 observations.
In this work, we utilize the NuSTAR Cycle 5+6+8+9 data and simultaneous XMM-Newton observations in order to perform the first spectroscopic analysis of the 60-source catalog.
We present this analysis and measure the $N_{{\mathrm{H}}}$ distribution of the sample. 
We measure an observed CT fraction of \mc{$0.13_{-0.04}^{+0.15}$} down to an observed $8-24$~keV flux of $6.0 \times 10^{-14}$~erg/s/cm$^{2}$, and we correct our analysis for absorption bias to estimate an underlying CT fraction of \mc{$0.32_{-0.08}^{+0.23}$}.
The derived obscuration distribution and CT fraction are consistent with population synthesis models and previous surveys.

\end{abstract}

\keywords{galaxy evolution: general; X-ray astronomy: general; galaxy clusters; general}

\section{INTRODUCTION}\label{sec:intro}

At the center of nearly every massive galaxy, we observe supermassive black holes (SMBHs) that are thought to co-evolve with their host galaxies \citep[e.g.][]{Ferrarese2000-AGN_n_gals, Gebhardt2000-SMBH_v_dv, Kormendy13}.
Accretion is the primary channel of black hole growth \citep[e.g.][]{Soltan82}, and during phases of high accretion, the SMBH is classified as an active galactic nucleus (AGN).
These accretion processes power AGN emission across the entire electromagnetic spectrum.
In the X-ray band, emission is produced by the corona: a reservoir of relativistic electrons located within a light hour of the SMBH \citep[e.g][]{Martocchia96-corona_location, Fabian09-corona}. 
X-rays are produced when optical and UV emission from the accretion disk is Comptonized by the corona \citep[e.g.][]{Haardt91-comptonization}, resulting in a powerlaw spectrum.
\mc{However, for many AGN, the intrinsic powerlaw emission is reprocessed via the photoelectric effect and Compton scattering by an obscuring region, which is known as the obscuring torus in typical models of AGN unification} \citep[e.g.][]{Turner97,Risaliti99-Seyfert2_Absorbtion,Brightman11-IRvX}.
The features of the observed X-ray spectrum are sensitive to the geometrical properties of the obscuring torus, particularly the line-of-sight column density ($N_{{\mathrm{H}}(los)}$; $N_{\mathrm{H}}$ for brevity). 
When $N_{{\mathrm{H}}} < 10^{22}$~cm$^{-2}$, the AGN is dubbed unabsorbed.
Between $10^{22}$~cm$^{-2} \leq N_{\mathrm{H}} < 10^{24}$~cm$^{-2}$, the AGN is classified as Compton Thin (C-Thin) or obscured.
In the most extreme case, when $N_{\mathrm{H}} \geq 1/\sigma_T \sim 10^{24}$~cm$^{-2}$, where $\sigma_T$ is the Thomson cross-section for electron scattering, the AGN is classified as Compton Thick (CT).
At these high column densities, the soft-band emission ($\lesssim 5$keV) is heavily suppressed, and the observed soft X-ray emission is 30-50 times fainter than the intrinsic X-ray luminosity \citep[$L_X$; see Figure 4 from][]{Aananna22}, making it difficult to identify the full CT population.
While hard X-rays ($\geq$ 10\,keV) are attenuated in the most extreme cases, they are less affected than the soft X-ray band \citep[e.g.][Figure 7]{Ueda14-CXB}, so deep exposures in this band are an excellent tool to study the properties of the full AGN population, including CT AGN.

\mc{Studying AGN above 10~keV also helps us understand the Cosmic X-ray Background (CXB), i.e. the diffuse X-ray emission covering the entire sky in the $\sim$1-300\,keV band. 
The vast majority of the CXB is unresolved emission of faint AGN. 
As a result, the CXB probes the full AGN population and has provided constraints for models of SMBH growth \citep[e.g.][]{Comastri95-CXB,Gilli07-CXB,Ueda14-CXB,Annana19-Accretion_HistoryI,Gerolymatou25}.
In particular, the peak of the CXB spectrum around $20-30$keV \citep[e.g,][]{Rossland23-neccessarily_lengthy} requires that a sizable fraction of AGN must be CT \citep[][]{Comastri95-CXB,Gilli07-CXB,Ueda14-CXB,Annana19-Accretion_HistoryI,Gerolymatou25}.
The CT fraction ($f_{CT}$, which we define as the fraction of all AGN with $N_{\mathrm{H}} > 10^{24}$~cm$^{-2}$; $N_\text{CT}/N_\text{total}$)} is an important component of SMBH population synthesis models, but measuring $f_{CT}$ from the CXB yields large uncertainties \citep{Gilli07-CXB,Treister09-CXB,Ueda14-CXB}.
In order to accurately constrain $f_{CT}$, it is necessary to resolve the AGN population that comprises the CXB.
AGN X-ray surveys are one of the best methods to accomplish this.

Previous surveys in the X-ray band have yielded constraints on $f_{CT}$.
\cite{Lansbury17-Serendipitous_40Month} finds $f_{CT}=0.3$ using the 40-month Serendipitous survey ($z \lesssim 0.1$). 
In the NuSTAR-COSMOS survey ($z \sim 0.5$), \cite{Civano15-NuSTAR_COSMOS} finds $f_{CT} = 0.13 \pm 0.03$ or $f_{CT} = 0.20 \pm 0.03$ based on a single CT detection.
Chandra observations of radio-selected galaxies at $0.5 < z < 1$ yielded $f_{CT}  \sim 0.20$ \citep{Kura21}, and a similar result was found for radio-selected galaxies at higher redshifts \citep[$1 < z < 2$;][]{Wilkes13}.
From the combined source catalogs of the NuSTAR extragalactic surveys, \cite{Zappacosta18-NUSTAR_extragal_hard-band} finds the CT fraction to be between $f_{CT} = 0.02-0.56$ ($z \sim 0.5$). From the Ultra Deep Field survey (UDF), \cite{Masini18-NuSTAR_UDF} finds $f_{CT} = 0.13 \pm 0.02$ ($z \sim 1$). 
\citet{TA21-CT} estimates that $\sim 8 \%$ of sources detected by the Swift-BAT survey ($z \leq 0.05$) are CT, while \citet{Akylas24-CT_local} estimates a local fraction $f_{CT} = 25-30\%$ by folding the BAT survey with mid-IR-selected AGN.
\citet{Carroll23} fuses joint Mid-IR and X-ray properties to find $f_{CT} = 0.555^{+0.040}_{-0.030}$.
Most recently, \citet{Boorman25} finds $f_{CT} = 0.35 \pm 0.09$ using local, NuSTAR-detected AGN selected by IR properties.

Several of these surveys utilize NuSTAR, which is the first X-ray telescope able to focus AGN emission at energies $>10$~keV.
The NuSTAR extragalactic surveys have followed a ``wedding cake'' strategy by combining relatively shallow observations of wide fields with deep, narrow surveys.
The wide layers of the wedding cake include the COSMOS Legacy survey \citep[1.7 deg$^{-2}$;][]{Civano15-NuSTAR_COSMOS} and the Serendipitous survey \citep[$13$ deg$^{-2}$;][]{Alexander13-NuSTAR_serendipitous,Lansbury17-Serendipitous_40Month}. 
The deep, narrow layers include the Extended Chandra Deep Field-South \citep[$\sim0.33$ deg$^{-2}$;][]{Mullaney15-ECDFS_NUSTAR} and the UDF \citep[$0.6$~deg$^{-2}$;][]{Masini18-NuSTAR_UDF}. 

The JWST-North Ecliptic Pole Time Domain Field (NEP; 0.3 deg$^{-2}$) is a recent addition to the wedding cake.
With 3.5 Ms of exposure time across five contiguous years, the NEP field hosts the deepest NuSTAR survey to date. 
The first results of the NuSTAR Cycle 5 data are presented by \cite{Zhao21_cycle5}, the Cycle 5+6 analysis is presented by \citet{Zhao24-cycle5-6} (\citetalias{Zhao24-cycle5-6} hereafter), and Cycle 8+9 is discussed in \cite{Silver25}. 
\citetalias{Zhao24-cycle5-6} identified 60 sources from the combined Cycle 5+6 NuSTAR observations. 
Hardness ratios---which give the relative difference between the 3-8 and 8-24~keV bands---yield predicted $N_{\mathrm{H}}$ values for the sources, and \citetalias{Zhao24-cycle5-6} uses these hardness ratios to estimate $f_{CT} = 0.18^{+0.20}_{-0.08}$. 
In this work, we follow up on the \citetalias{Zhao24-cycle5-6} catalog by presenting the first NuSTAR+XMM-Newton spectral analysis of these 60 sources, and we calculate the obscuration distribution and $f_{CT}$ of the sample.

In Section \ref{sec:DR}, we describe the sample, observations, and data reduction. 
In Section \ref{sec:analysis}, we define the models used to analyze our X-ray spectra and explain our fitting methods. 
We calculate the $N_{\mathrm{H}}$ distribution in Section \ref{sec:results} and discuss the implications in Section \ref{sec:discussion}. Our findings are summarized in Section \ref{sec:conclusion}.
This study assumes a $\Lambda$CDM cosmology with {\it H$_{0}$} = 70 km s$^{-1}$ Mpc$^{-1}$, $\Omega_{M}$ = 0.27, and $\Omega_{\Lambda}$ = 0.73 \citep{Planck20}.

\section{SAMPLE AND DATA REDUCTION}\label{sec:DR}

\begin{table*}[t]
    \centering
    \caption{NuSTAR and XMM-Newton Observations of the NEP\label{tab:obs}}
    \begin{tabular}{ccc|c|c}
     \hline 
     \hline
     \\[-0.75em]
     \textbf{Cycle} & \textbf{Epoch} & \textbf{Date} &  \textbf{NuSTAR IDs} & \textbf{XMM-Newton ID} \\
         \hline
 5 & 1 & 09/2019 & 60511001002, 60511002002, 60511003002 & - \\
   & 2 & 01/2020 & 60511004002, 60511005002, 60511006002 & - \\
   & 3 & 03/2020 & 60511007002, 60511008001, 60511009001 & - \\
 \hline
 6 & 4 & 10/2020 & 60666001002, 60666002002, 60666003002 & 0870860101 \\
   & 5 & 01/2021 & 60666004002, 60666005002, 60666006002 & 0870860201 \\
   & 6 & 10/2021 & 60666007002, 60666008002, 60666009002 & - \\
   & 7 & 01/2022 & 60666010002, 60666011002, 60666012002 & 0870860401 \\
 \hline
 8 & 8 & 08/2022 & 60666013002, 60666014002, 60666015002 & 0870860501 \\
   & 9 & 11/2022 & 60810001002, 60810002002, 60810003002 & 0913590101  \\
   & 10& 02/2023 & 60810004002, 60810005002, 60810006002 &  0913590501\\
   & 11& 05/2023 & 60810007002, 60810008002, 60810009002, 60810009004 & 0913590601 \\
  \hline
 9 & 12& 08/2023 & 60910001002, 60910002002, 60910003002, 60910001004 & 0931420701 \\
   & 13& 11/2023 & 60910004002, 60910005002, 60910006002 & 0931420101 \\
   & 14& 05/2024 & 60910007002, 60910008002, 60910009002 & 0931420101 \\
   & 15& 02/2024 & 60910010002, 60910011002, 60910012002 & 0931420501 \\
         \\[-0.5em]
      \hline
    \end{tabular}
\end{table*}

The observation IDs used in this analysis are listed in Table \ref{tab:obs}, and the sources from the \citetalias{Zhao24-cycle5-6} catalog are summarized in Table \ref{tab:agn_props}.
The following procedure uses all available NuSTAR data (Cycles 5+6+8+9) and XMM-Newton observations in order to extract spectra for these 60 sources.

\subsection{NuSTAR}

The observations of the NuSTAR-NEP survey were taken in fifteen epochs, with three or four observations per epoch (Table \ref{tab:obs}). These epochs took place during NuSTAR Cycle 5 (PI: Civano, ID: 5192), Cycle 6 (PI: Civano, ID: 6218), Cycle 8 (PI: Civano, ID: 8180), and Cycle 9 (PI: Civano, ID: 9267). 
\mc{Data reduction was performed by \cite{Zhao21_cycle5} (Cycle 5), \citetalias{Zhao24-cycle5-6} (Cycle 6), and \cite{Silver25} (Cycles 8 and 9). We discuss relevant details here.} 

\mc{The NuSTAR data were processed using HEASOFT (v.27.2, v.6.29c, and v.6.33.1) and CALDB (v.202005266, v.20211115, and v.20230816). 
Data from all observations were calibrated, cleaned, and screened using \textsc{nupipeline}.
Time intervals with high background were removed by screening for periods in which the 3.5-9.5~keV count rate was $\geq 2$ times higher than the average value within the entire observation.}

\mc{Vignetted exposure maps were produced using the \textsc{nuexpomap} tool from NuSTARDAS. 
Background maps were created using \textsc{nuskybgd}\footnote{https://github.com/NuSTAR/nuskybgd} \citep{Wik14-nuskybgd}. 
We refer to \cite{Zhao21_cycle5}, \citetalias{Zhao24-cycle5-6}, and \cite{Silver25} for further details.}

For spectral extraction, region sizes were chosen to maximize individual source signal-to-noise (see Appendix \ref{sec:a_r} for details), and \textsc{nuproducts} was used to extract higher level data products from the exposure-corrected NuSTAR data.
Background spectra were extracted using the \textsc{nubgdspec} routine from \textsc{nuskybgd}  \citep{Wik14-nuskybgd}.
In instances where background-subtracted spectra had net zero or negative source photons---implying that the spectrum is background-dominated---that observation was excluded from the analysis of the source.

Finally, the spectra from all observations were added using the \textsc{addspec} python routine\footnote{\url{https://github.com/JohannesBuchner/addspec.py}}. 
FPMA and FPMB were kept separate. 
In this paper, we assume that all the sources have zero variability and combine spectra from all epochs to improve count statistics.
\citetalias{Zhao24-cycle5-6} discusses variability analysis.

\subsection{XMM-Newton}

XMM-Newton observations were timed such that they aligned with the NuSTAR epochs in Cycles 6, 8, and 9 (see Table \ref{tab:obs}). 
One XMM-Newton observation (Obs. ID 0870860301; NuSTAR epoch 6) was background dominated and not included in any analysis.
Otherwise, each epoch in NuSTAR cycles 6, 8, and 9 have a corresponding XMM-Newton observation, which were reduced by \citetalias{Zhao24-cycle5-6} (Cycle 6) and \cite{Silver25} (Cycles 8 and 9) using the XMM-Newton Science Analysis System (SAS; version 20.0.0) \mc{and following the procedures outlined by  \cite{Brunner08_LockmanHole}, \cite{Cappelluti09-XMM_COSMOS_PointCat}, and \cite{LaMassa09_XMM_Seyfert}.
The tasks \textsc{emproc} and \textsc{epproc} were used to create observational data files of the three instruments (MOS1, MOS2, and PN), from which high background time intervals and fluorescent emission line energy bands were excluded. 
From the clean event files, images were generated for all three instruments in the 0.5-2~keV and 2-10~keV energy bands.
For both of these energy bands, exposure maps were generated using SAS \textsc{eexpmap} task.
After masking potential X-ray sources identified by \textsc{eboxdetect}, background maps were generated for all three instruments using \textsc{esplinemap}, with \textsc{fitmethod} = model. 
This models the XMM-Newton background with a detector component and a CXB component.}

Out of the 60 NuSTAR-identified sources in the \citetalias{Zhao24-cycle5-6} catalog, 36 have lower-energy XMM-Newton counterparts.
XMM-Newton spectra were extracted using $15\arcsec$ source regions centered around the coordinates of the XMM-Newton sources, and local backgrounds were extracted from 75-100$\arcsec$ annuli. 
XMM-Newton has a smaller PSF (FWHM $5-6 \arcsec$) and does not require the SNR-maximized region method used for NuSTAR.
The \textsc{evselect} routine was used to extract source and background spectra from MOS1, MOS2, and PN.
The scaling factor for normalizing the background spectrum was calculated using the \textsc{backscale} routine.
The \textsc{rmfgen} and \textsc{arfgen} tasks generated RMFs and ARFs, respectively.
In order to account for the cross calibration between NuSTAR and XMM-Newton instruments, we applied the flags \textsc{applyabsfluxcorr=yes} (which corrects the effective area of PN spectra in order to better match NuSTAR) and \textsc{applyxcaladjustment=yes} (which improves consistency between MOS and PN spectra by applying an energy-dependent correction function) to the \textsc{arfgen} routine.

\startlongtable
\tabletypesize{\tiny}
\begin{deluxetable*}{cccc|ccccccccccc}
\tablecaption{AGN Properties\label{tab:agn_props}}
\tiny
\startdata
\\
 Nu & XMM & Nu &  XMM & RA & DEC & $z$\tablenotemark{c} & $F^{2-10}_{obs}$  & $L^{2-10}_{obs}$ &  $F^{3-24}_{obs}$  & $L^{3-24}_{obs}$ & $F^{8-24}_{obs}$ & $L^{8-24}_{obs}$  &    \\
 ID\tablenotemark{a} & ID\tablenotemark{a}\tablenotemark{b} & net cts & net cts & [deg] & [deg]  &  &  [erg/s/cm$^2$] & [erg/s] &  [erg/s/cm$^2$] & [erg/s] & [erg/s/cm$^2$] & [erg/s] &  Class\tablenotemark{d}    \\
 \hline
1 & - & 37.2 & - & 260.530704 & 66.059882 & - & - & - & 1.69e-13 & - & 2.08e-13 & - & - \\
2 & 4 & 476.7 & 1148.9 & 261.004963 & 66.012388 & 0.892$^\text{s}$ & 6.01e-14 & 2.37E+44 & 1.07e-13 & 4.21E+44 & 1.3e-13 & 5.12E+44 & Q (Z24) \\
3 & 60 & 183.4 & 66.1 & 260.615623 & 65.980155 & 0.0$^\text{s}$ & 1.23e-14 & - & 2.8e-14 & - & 4.27e-14 & - & S (Z24) \\
4 & - & 424.4 & - & 260.74987 & 65.968707 & - & - & - & 2.35e-14 & - & 3.24e-14 & - & - \\
5 & - & 149.9 & - & 260.434776 & 65.961318 & - & - & - & 2.9e-14 & - & 3.85e-14 & - & - \\
6 & 37 & 575.0 & 200.3 & 260.522038 & 65.954027 & 0.62$^\text{p}$ & 2.66e-14 & 4.30E+43 & 7.39e-14 & 1.19E+44 & 4.98e-14 & 8.04E+43 & Q (O26) \\
7 & 34 & 304.9 & 94.6 & 260.468713 & 65.93478 & 0.27$^\text{s}$ & 9.43e-15 & 2.14E+42 & 2.28e-14 & 5.16E+42 & 3.5e-14 & 7.93E+42 & G (Z24) \\
8 & 97 & 160.3 & 30.8 & 260.622873 & 65.923493 & 0.81$^\text{p}$ & 8.72e-15 & 2.71E+43 & 1.29e-14 & 4.01E+43 & 2.1e-14 & 6.52E+43 & Q (O26) \\
9 & - & 114.6 & - & 261.188068 & 65.881698 & - & - & - & 7.26e-14 & - & 1.1e-13 & - & - \\
10 & 240 & 188.3 & 48.1 & 260.544655 & 65.887843 & - & 4.48e-15 & - & 1.29e-14 & - & 2.02e-14 & - & - \\
11 & 32/138 & 231.0 & 246.7 & 260.73184 & 65.89017 & 0.601$^\text{s}$ & 9.33e-15 & 1.40E+43 & 1.26e-14 & 1.89E+43 & 1.76e-14 & 2.63E+43 & Q (Z24) \\
12 & 26 & 322.9 & 106.1 & 260.466229 & 65.879465 & 5.35$^\text{p}$ & 7.98e-15 & 2.44E+45 & 2.26e-14 & 6.90E+45 & 3.2e-14 & 9.76E+45 & Q (O26) \\
13 & 63/76 & 368.5 & 188.4 & 260.783715 & 65.872453 & - & 4.74e-15 & - & 1.38e-14 & - & 1.93e-14 & - & - \\
14 & - & 278.1 & - & 260.359797 & 65.853886 & - & - & - & 2.89e-14 & - & 4.34e-14 & - & - \\
15 & 56 & 140.4 & 60.4 & 260.539214 & 65.790189 & 2.251$^\text{s}$ & 8.73e-15 & 3.37E+44 & 1.57e-14 & 6.05E+44 & 1.98e-14 & 7.64E+44 & Q (Z24) \\
16 & - & 216.5 & - & 261.103491 & 65.770554 & - & - & - & 5.21e-14 & - & 6.81e-14 & - & - \\
17 & - & 254.0 & - & 260.978502 & 65.723582 & - & - & - & 3.53e-14 & - & 3.29e-14 & - & - \\
18 & - & 146.2 & - & 260.403922 & 65.720876 & - & - & - & 4.58e-14 & - & 7.35e-14 & - & - \\
19 & 49 & 855.5 & 207.2 & 260.689412 & 65.740591 & 1.32$^\text{p}$ & 1.85e-14 & 1.92E+44 & 3.7e-14 & 3.84E+44 & 3.45e-14 & 3.58E+44 & Q (O26) \\
20 & 11 & 158.5 & 220.3 & 260.867239 & 65.982407 & 1.425$^\text{s}$ & 1.58e-14 & 1.98E+44 & 7.15e-14 & 8.97E+44 & 7.33e-14 & 9.20E+44 & Q (Z24) \\
21 & 12 & 408.9 & 607.3 & 260.823684 & 65.944847 & 0.495$^\text{s}$ & 2.47e-14 & 2.32E+43 & 3.96e-14 & 3.71E+43 & 2.84e-14 & 2.66E+43 & G (Z24) \\
22 & 47 & 382.2 & 123.4 & 260.656285 & 65.931053 & 1.23$^\text{p}$ & 1.44e-14 & 1.26E+44 & 4.89e-14 & 4.26E+44 & 2.8e-14 & 2.44E+44 & Q (O26) \\
23 & 112/165 & 214.3 & 117.2 & 260.495899 & 65.907535 & - & 6.11e-15 & - & 2.1e-14 & - & 7.65e-15 & - & - \\
24 & 3 & 347.7 & 1451.7 & 260.42374 & 65.921001 & 0.523$^\text{s}$ & 6.11e-14 & 6.54E+43 & 8.88e-14 & 9.50E+43 & 4.4e-14 & 4.71E+43 & Q (Z24) \\
25 & - & 196.0 & - & 261.194464 & 65.859152 & - & - & - & 1.03e-13 & - & 8.76e-14 & - & - \\
26 & 36 & 243.6 & 165.9 & 261.025736 & 65.842349 & 1.8$^\text{p}$ & 2.1e-14 & 4.69E+44 & 2.15e-14 & 4.80E+44 & 1.8e-14 & 4.02E+44 & Q (O26) \\
27 & - & 414.6 & - & 260.523638 & 65.834538 & - & - & - & 1.37e-14 & - & 1.17e-14 & - & - \\
28 & 7 & 557.8 & 495.3 & 260.965651 & 65.836937 & 0.673$^\text{s}$ & 2.01e-14 & 3.96E+43 & 3.15e-14 & 6.21E+43 & 1.48e-14 & 2.92E+43 & Q (Z24) \\
29 & 1 & 8447.9 & 7746.7 & 260.808941 & 65.796673 & 1.441$^\text{s}$ & 1.51e-13 & 1.95E+45 & 2.79e-13 & 3.60E+45 & 1.65e-13 & 2.13E+45 & Q (Z24) \\
30 & 73 & 299.4 & 77.7 & 260.44349 & 65.839801 & 1.68$^\text{p}$ & 1.5e-14 & 2.82E+44 & 1.62e-14 & 3.05E+44 & 1.5e-14 & 2.82E+44 & Q (O26) \\
31 & 51/64 & 411.1 & 85.0 & 261.078906 & 65.815655 & 0.376$^\text{s}$ & 2.13e-14 & 1.04E+43 & 4.89e-14 & 2.39E+43 & 2.39e-14 & 1.17E+43 & G (Z24) \\
32 & 69 & 334.1 & 122.2 & 260.695612 & 65.808196 & 1.83$^\text{p}$ & 5.69e-15 & 1.32E+44 & 1.47e-14 & 3.42E+44 & 7.13e-15 & 1.66E+44 & Q (O26) \\
33 & - & 126.1 & - & 261.11552 & 65.810138 & - & - & - & 6.96e-14 & - & 5.17e-14 & - & - \\
34 & 169 & 181.0 & 128.5 & 260.885542 & 65.781661 & 0.403$^\text{s}$ & 6.49e-15 & 3.73E+42 & 1.12e-14 & 6.44E+42 & 9.14e-15 & 5.25E+42 & G (Z24) \\
35 & - & 146.2 & - & 260.53465 & 65.756721 & - & - & - & 2.68e-14 & - & 1.54e-14 & - & - \\
36 & 66 & 457.4 & 68.2 & 260.58976 & 65.734981 & 1.019$^\text{s}$ & 1.18e-14 & 6.46E+43 & 1.79e-14 & 9.80E+43 & 2.23e-14 & 1.22E+44 & Q (Z24) \\
37 & 205 & 299.5 & 74.1 & 261.035039 & 65.734455 & 0.85$^\text{p}$ & 2.17e-14 & 7.59E+43 & 3.49e-14 & 1.22E+44 & 3.37e-14 & 1.18E+44 & Q (O26) \\
38 & - & 415.5 & - & 261.006258 & 65.635478 & - & - & - & 2.56e-13 & - & 2.14e-13 & - & - \\
39 & 99 & 124.3 & 19.9 & 260.70263 & 66.014412 & 4.35$^\text{p}$ & 9.99e-15 & 1.87E+45 & 7.78e-14 & 1.46E+46 & 1.07e-13 & 2.01E+46 & Q (O26) \\
40 & - & 161.9 & - & 260.99109 & 65.922953 & - & - & - & 3.86e-14 & - & 3.72e-14 & - & - \\
41 & 10/203 & 346.0 & 470.6 & 260.751816 & 65.908622 & - & 1.29e-14 & - & 2.3e-14 & - & 1.93e-14 & - & - \\
42 & - & 122.6 & - & 261.054304 & 65.870312 & - & - & - & 3.03e-14 & - & 3.11e-14 & - & - \\
43 & 5 & 684.6 & 1569.3 & 260.638622 & 65.84636 & 1.337$^\text{s}$ & 2.22e-14 & 2.38E+44 & 2.35e-14 & 2.52E+44 & 1.48e-14 & 1.59E+44 & Q (Z24) \\
44 & - & 312.0 & - & 260.718727 & 65.845092 & - & - & - & 1.5e-14 & - & 1.45e-14 & - & - \\
45 & 141 & 482.6 & 94.0 & 260.990613 & 65.824629 & 1.35$^\text{p}$ & 7.99e-15 & 8.77E+43 & 4.52e-14 & 4.96E+44 & 3.05e-14 & 3.35E+44 & Q (O26) \\
46 & 54/65 & 254.6 & 0.0 & 260.769055 & 65.77412 & 0.634$^\text{s}$ & 8.06e-15 & 1.37E+43 & 3.05e-14 & 5.20E+43 & 2.27e-14 & 3.87E+43 & Q (Z24) \\
47 & - & 396.8 & - & 260.767207 & 65.821916 & - & - & - & 2.15e-14 & - & 1.42e-14 & - & - \\
48 & - & 262.0 & - & 260.673401 & 65.659318 & - & - & - & 6.31e-14 & - & 1.01e-13 & - & - \\
49 & - & 199.7 & - & 260.982765 & 65.932518 & - & - & - & 3.86e-14 & - & 5.12e-14 & - & - \\
50 & - & 240.6 & - & 260.801946 & 65.929818 & - & - & - & 1.56e-14 & - & 2.03e-14 & - & - \\
51 & 134/181 & 227.9 & 14.7 & 260.557293 & 65.948639 & 0.47$^\text{p}$ & 8.27e-15 & 6.85E+42 & 3.46e-14 & 2.86E+43 & 4.92e-14 & 4.07E+43 & N/A (Z24) \\
52 & - & 227.9 & - & 260.823766 & 65.895685 & - & - & - & 1.44e-14 & - & 2.42e-14 & - & - \\
53 & 13 & 983.6 & 256.5 & 260.778758 & 65.856746 & 0.885$^\text{s}$ & 2.73e-14 & 1.05E+44 & 6.85e-14 & 2.65E+44 & 4.76e-14 & 1.84E+44 & Q (Z24) \\
54 & 14 & 547.4 & 326.5 & 260.893905 & 65.819897 & 0.779$^\text{s}$ & 2.41e-14 & 6.80E+43 & 1.62e-14 & 4.57E+43 & 2.2e-14 & 6.21E+43 & G (Z24) \\
55 & 2 & 1778.1 & 2622.3 & 260.429729 & 65.766017 & 0.781$^\text{s}$ & 1.01e-13 & 2.87E+44 & 1.59e-13 & 4.52E+44 & 9.55e-14 & 2.71E+44 & Q (Z24) \\
56 & - & 242.3 & - & 260.566157 & 65.764297 & - & - & - & 1.56e-14 & - & 3.25e-14 & - & - \\
57 & 28 & 359.4 & 131.1 & 260.493365 & 65.736825 & 1.349$^\text{s}$ & 1.12e-14 & 1.23E+44 & 3.28e-14 & 3.59E+44 & 3.11e-14 & 3.41E+44 & Q (O26) \\
58 & 17 & 2586.7 & 577.1 & 260.671348 & 65.711891 & 0.179$^\text{s}$ & 5.16e-14 & 4.63E+42 & 1.98e-13 & 1.78E+43 & 1.44e-13 & 1.29E+43 & G (Z24) \\
59 & - & 178.8 & - & 260.807485 & 65.705176 & - & - & - & 3.55e-14 & - & 5.22e-14 & - & - \\
60 & 42 & 255.8 & 0.0 & 260.726199 & 65.698978 & 0.44$^\text{p}$ & 1.11e-14 & 7.85E+42 & 3.08e-14 & 2.18E+43 & 5.02e-14 & 3.55E+43 & N/A (Z24) \\
\enddata
\tablecomments{ \\
$^{a}$ NuSTAR and XMM-Newton IDs from the \citetalias{Zhao24-cycle5-6} catalog \\
$^{b}$ For sources with more than one XMM-Newton match, both are listed. The first match is used for the analysis. \\
$^{c}$ superscript denotes method for redshift measurment (p for photometric, s for spectroscopic) \\
$^{d}$ S for Star, Q for Quasar, G for Galaxy, N/A for unidentified. In parentheses, the work that presents the classification (\citetalias{Zhao24-cycle5-6} or O26) is cited. Table \ref{tab:SED} gives SED inferences for sources classified by O26.\\\
}
\end{deluxetable*}

\subsection{Multi-Wavelength Counterparts}\label{sec:samp}

\begin{figure*}
    \centering
    \includegraphics[width=0.8\textwidth]{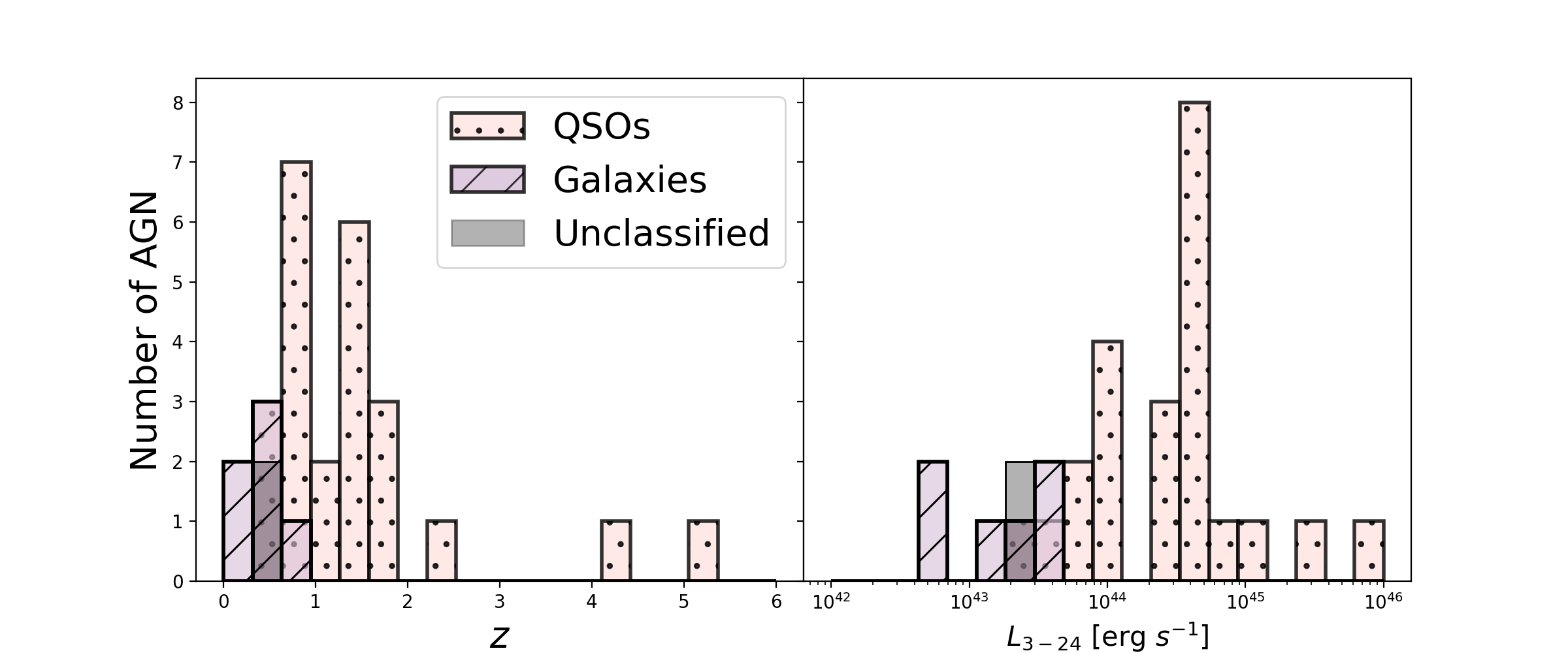}
    \caption{\mc{The redshifts (left) and observed} 3-24~keV X-ray luminosities (right) for sources with detected multi-wavelength counterparts. Shown are quasars (dotted; pink), galaxies (hashed; purple), and unclassified (solid; gray) objects.}
    \label{fig:z_L}
\end{figure*}

Of the 60 sources in the \citetalias{Zhao24-cycle5-6} catalog, 33 have multi-wavelength matches, 
\mc{which allow redshifts and classification to be inferred. 
The redshift and luminosity distributions for these 33 sources are shown in Figure \ref{fig:z_L}.}

\mc{\citetalias{Zhao24-cycle5-6} identified 21 multi-wavelength counterparts, which provide spectroscopic redshifts. 
These sources are identified as quasars (which have broad emission lines), composite galaxies (including Type II AGN and star forming galaxies with narrow emission lines), or stars.
\citetalias{Zhao24-cycle5-6} classifies objects by cross-correlating optical counterparts against SDSS template spectra\footnote{https://classic.sdss.org/dr5/algorithms/spectemplates/}.}

\mc{In addition to the 21 counterparts identified by \citetalias{Zhao24-cycle5-6}, Ortiz et al. \textit{in prep} (O26 hereafter) cross-matches eleven sources with recent catalogs.
One source (AGN 57) is matched with the DESI DR1 catalog \citep{DESI_D1} and has a corresponding spectroscopic redshift.
Eleven sources have counterparts in the JWST and HST footprints (O26).
Photometric redshifts and galaxy types are inferred using SED inferences from X-CIGALE \citep{Boquien19-CIGALE,Yang20-CIGALE} using all available broadband photometry from space (JWST \citep[][Jansen et al \textit{in prep}]{Windhorst23-NEP}); HST \citep{OBrien24}; WISE \citep{unWISE19}) and the ground (HSC \citep{HSC18}, SDSS \citep{SDSS_DR17}, MMT \citep{Willmer23-MMT}, JPAS \citep{Hernan-Caballero23-JWST_NEP_z}). 
The parameters of interest from SED fitting are the photometric redshift quality parameter $Q_z$ \citep[see equation 8 of][]{Brammer08} and $f_{AGN}$ (fractional contribution of the AGN SED to the 0.1-30 $\mu$m bolometric luminosity), which we use to classify objects.
Sources with $f_{AGN} > 0.2$ are generally inferred to be QSOs, but due to large uncertainties \citep[$\sim 5$-10$\%$;][]{Ciesla15}, we require $f_{AGN} > 0.3$ for QSO classification; otherwise, the source is classified as a composite galaxy.
These relevant details are summarized in Table \ref{tab:SED}.
Full details on the catalog creation and SED fitting are to be discussed in O26.}



\begin{table}[t]
    \centering
    \caption{O26 SED inferences \label{tab:SED}}
    \begin{tabular}{ccccc}
     \hline
     \hline 
     \\[-0.75em]
      Nu ID & $z$ & $f_{AGN}$\footnote{fraction of emission attributed to AGN} & Class\footnote{Q for Quasar, G for Galaxy} & $Q_z$\footnote{Redshift quality parameter \citep{Brammer08}, with $Q_z < 1$ deemed reliable} \\\\[-0.75em]
     	\hline
	\hline
6 & 0.62 & 0.233 & G & 0.07 \\
8 & 0.81 & 0.414 & Q & 0.14 \\
12 & 5.35 & 0.233 & G & 2.94 \\
19 & 1.32 & 0.527 & Q & 0.8 \\
22 & 1.23 & 0.498 & Q & 0.43 \\
26 & 1.8 & 0.441 & Q & 1.17 \\
30 & 1.68 & 0.569 & Q & 2.06 \\
32 & 1.83 & 0.303 & Q & 1.97 \\
37 & 0.85 & 0.382 & Q & 0.31 \\
39 & 4.35 & 0.201 & G & 1.20 \\
45 & 1.35 & 0.345 & Q & 0.07 \\
57 & 1.35 & 0.384 & Q & -\footnote{AGN 57 has a spectroscopic redshift from \cite{DESI_D1}, so $Q_z$ is not listed} \\
      \hline
    \end{tabular}
\end{table}

\section{SPECTRAL ANALYSIS}\label{sec:analysis}
\begin{figure}
    \centering
    \includegraphics[width=0.49\textwidth]{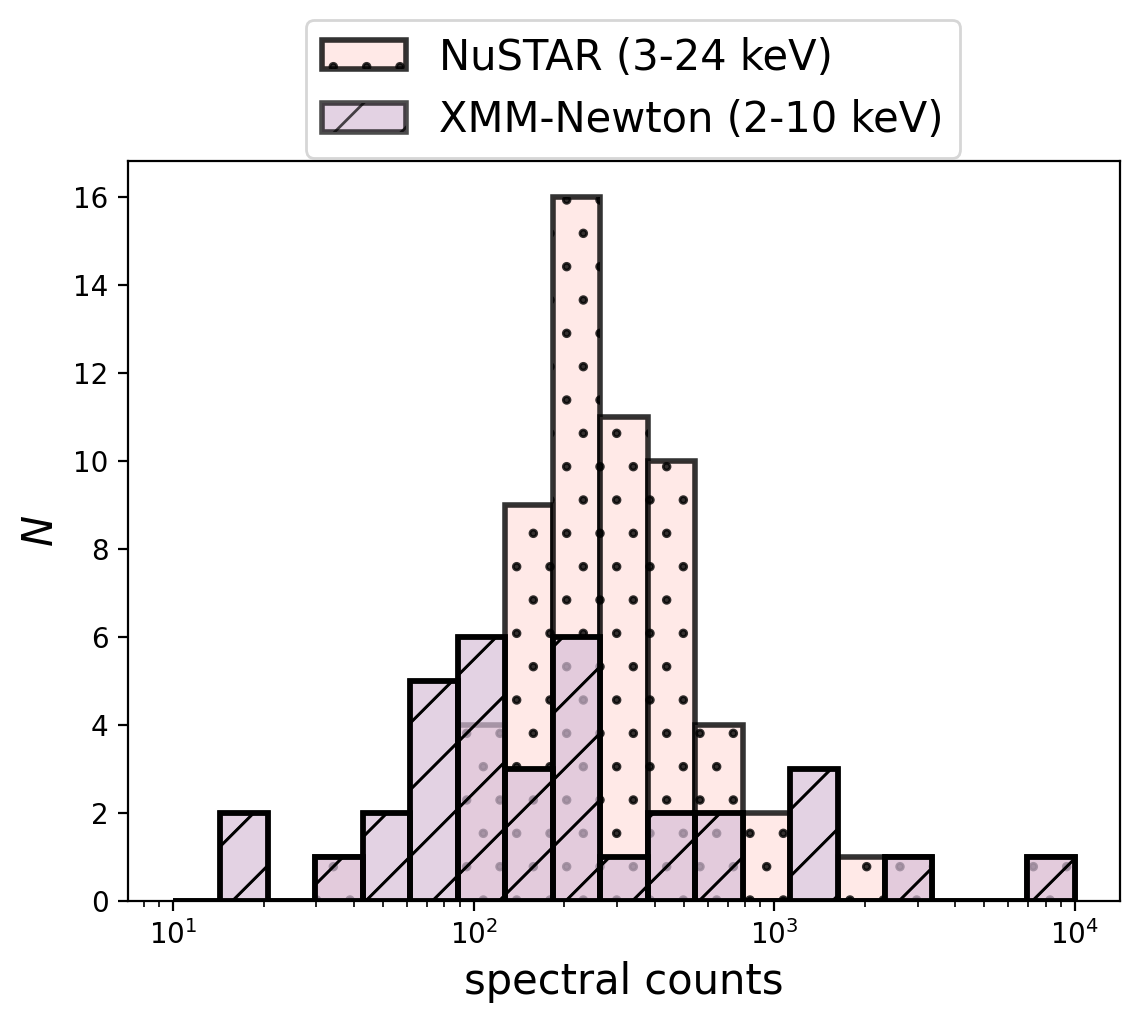}
    \caption{A histogram of the net spectral counts in NuSTAR (pink; dotted) and XMM-Newton (purple;  hashed) for the sources in our sample.}
    \label{fig:spec_cts}
\end{figure}

Together, NuSTAR and XMM-Newton spectra cover a $0.1 - 24.0$~keV energy range. 
Figure \ref{fig:spec_cts} shows the net spectral counts. 
Of the 36 sources with XMM-Newton counterparts, 27 have more NuSTAR counts than XMM-Newton counts; this is because most of the NuSTAR source regions are larger than the $15 ''$ XMM-Newton regions (Appendix \ref{sec:a_r}).
For spectral fitting, the count differences are corrected by the effective area encoded in the ARF files.


\subsection{Components of AGN X-ray Spectra}
\label{sec:comp}

\begin{figure}
    \centering
    \includegraphics[width=0.49\textwidth]{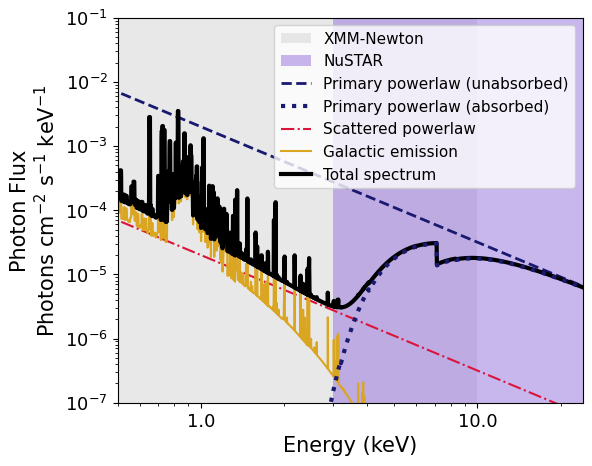}
    \caption{A schematic demonstrating the spectral components (unabsorbed (purple; dashed) and absorbed (purple; dotted) primary powerlaw, scattered powerlaw (red; dot-dashed), and galactic emission (gold; solid)) of the baseline model described in Section \ref{sec:models}. This model is shown for a typical Compton Thin AGN with $N_{\mathrm{H}} = 10^{23}$~cm$^{-2}$. The total spectrum is shown in bolded black. The background is shaded to demonstrate the energy range of XMM-Newton (gray) and NuSTAR (purple).}
    \label{fig:spec_schem}
\end{figure}

In this work, X-ray spectra are described using three additive components: a transmitted powerlaw, a scattered powerlaw, and galactic emission. 
Figure \ref{fig:spec_schem} shows a schematic of these components, and they are described in further detail below.

The transmitted powerlaw originates from the corona, which up-scatters UV and optical emission from the accretion disk into the X-ray band. 
Obscuring material along the line of sight suppresses soft X-rays, creating a low-energy cutoff whose location depends on $N_{\mathrm{H}}$.

The scattered powerlaw accounts for coronal emission that is Thomson scattered by warm photoionized material in the circumnuclear region \citep[][]{Matt19-hot_mirror}.
The parameters for the transmitted and scattered emission are equivalent, and the strength of the scattered component is controlled by a constant ($f_{\textsc{scatt}}$) that is free to vary between $10^{-5}$ and $0.1$ \citep{Buchner19_clumpy,  Gupta21}.

A \textsc{mekal} component \citep{mekal1,mekal2,mekal3} is commonly used to account for the $<3$~keV emission of hot, diffuse gas within the host galaxy \citep[e.g,][]{Mineo12,TA21-CT,Silver22,Diaz23,TA23}.
We refer to this as the galactic component. 
We keep all parameters frozen except for the normalization ($n_{\text{gal}}$), fixing the temperature to $kT = 0.6$~keV \citep{Gierlinski04-soft}, the hydrogen density to $1$~cm$^{-3}$, and abundance to solar \citep[1.0;][]{Crummy06}. The redshift is that of the host galaxy, and the switch parameter is set to 1, meaning that the spectrum is interpolated from a pre-existing table.

Many AGN models also include a reflected component, which models X-rays that are reprocessed by circumnuclear material in the accretion disk and torus. 
This component creates a ``Compton hump'' at $\sim 30$~keV (rest frame) for obscured AGN and has negligible effects below those energies \citep[see Figure 3 of][]{Carroll23} aside from an Fe K$\alpha$ florescence line.
All data are ignored above $24$~keV and the SNR of the X-ray spectra in this work are generally too low to detect the Fe K$\alpha$ line, so a separate reflected component would not be well constrained.
However, one of our two models (Section \ref{sec:models}) self-consistently includes reflected emission without requiring additional free parameters.

\subsection{The Models}
\label{sec:models}

\mc{We employ two models for our fitting: a simplified ``baseline" model that represents AGN as an obscured powerlaw, and a physically motivated torus model that more accurately reproduces the spectra of obscured AGN.
While the baseline model is sufficient for unobscured sources, torus models self-consistently account for photoelectric absorption alongside Compton scattering, cold reflection, and florescent emission, all of which are negligible for unabsorbed AGN but become important at log$(N_H) \gtrsim 23$ \citep{Yaqoob12_MYTorus, Balokovic18_borus, Buchner19_clumpy}.
Therefore, the torus models gives a more accurate representation of the parameter space for these C-Thin and CT sources.
When interpreting results (Section \ref{sec:results}), we follow \citet{Zappacosta18-NUSTAR_extragal_hard-band} by employing the baseline model for sources with log$(N_H) < 23$ and the clumpy torus model for log$(N_H) \geq 23$.}

\mc{The baseline model is defined as follows:}

\begin{equation}
 \begin{split}
    \textsc{phabs}_{\textsc{Gal}} *( \textsc{zphabs}_{\textsc{tor}}*\textsc{zpowerlw}~ + \\
    f_{\textsc{scatt}} * \textsc{zpowerlw} + \textsc{mekal).}
\end{split}
\end{equation}

The baseline model has two absorption components (\textsc{phabs}): one for Galactic absorption from the Milky Way \citep{H_map90,H_map05,H_map16}, and another representing intrinsic absorption from the AGN torus or host galaxy ($N_{\mathrm{H}}$).
The corona is modeled by a redshifted powerlaw (\textsc{zpowerlw}), which has two free parameters: the photon index ($\Gamma$ hereafter) and the normalization.
The strength of the scattered powerlaw is controlled by a constant ($f_{\textsc{scatt}}$).
The galactic emission is represented by \textsc{mekal}.

The second model used to describe the X-ray spectra employs the \textsc{uxclumpy} torus model by \cite{Buchner19_clumpy}, which is widely used for analyzing obscured AGN \citep[i.e.][]{Marchesi22-NGC1358,TA23,Akylas24-CT_local,Boorman25,TA25-Mrk477}. 
This model follows the formalism of the \textsc{Clumpy} model for IR emission \citet{Nenkova08-clumpy1,Nenkova08-clumpy2}, which describes the dusty torus as a gaussian distribution of clouds with various column densities.
\textsc{uxclumpy} was designed to match the frequency of eclipsing events observed by \citet{Markowitz14-eclipse} and the $N_{\mathrm{H}}$ distribution measured by \citet{Aird15}, \citet{Buchner15-PopSynth}, and \citet{Ricci15}.
We hereafter refer to this as the clumpy model:

\begin{equation}
 \begin{split}
 \textsc{phabs} *( \textsc{uxclumpy-cutoff}  + f_{\textsc{scatt}} * \\
 \textsc{uxclumpy-cutoff-omni} + \textsc{mekal).}
\end{split}
\end{equation}

In the clumpy model, the \textsc{phabs} component represents absorption from the Milky Way.
Absorption from obscuring material within the circumnuclear region ($N_{\mathrm{H}}$) is a free parameter of \textsc{uxclumpy-cutoff}, which self-consistently models the transmitted and reflected X-ray emission from a clumpy torus surrounding the central engine. \textsc{uxclumpy-cutoff-omni} represents the scattered component, which is modeled as the transmitted powerlaw scattered by warm material beyond the influence of the obscuring torus.
Just as with the baseline model, the strength of the scattered component is controlled by a constant $f_{\textsc{scatt}}.$ Galactic emission is represented by a \textsc{mekal} component.

As is standard for analysis of faint sources, torus parameters are kept fixed to average values. The only \textsc{uxclumpy} parameters kept free are $N_{\mathrm{H}}$, $\Gamma$, and the powerlaw normalization.
The high energy cutoff is set to $400$~keV \citep{Balokovic20-Ecut}. The parameters TORSigma and CTKcover---which together describe the covering fraction of the obscuring material---have minor effects on the spectra in the 0.1-24~keV energy range, so we fix them to median values ($45$ and $0.4$, respectively). 
The inclination angle ($\Theta_{\text{inc}}$) is set to $90.0 \degree$, which represents an edge-on AGN. 
Following unification schemes \citep{Antonucci93, Urry95-Unification, Ricci17-BAT}, unobscured AGN have lower $\Theta_{\text{inc}}$.
However, this parameter has negligible effects on the fits, so we keep it fixed throughout the process.

For both models, the redshift is that of the host galaxy. Sources without multi-wavelength counterparts are assigned $z=0.5$, which is the median redshift of the sample.

\subsection{Fitting Procedure}
\label{sec:BXA}


To begin the fitting procedure, we first determined whether to employ scattered powerlaw or galactic emission components.
If there are no XMM-Newton data for a source, both components are excluded.
Where XMM-Newton spectra exist, low signal-to-noise in the soft end leads to degeneracy between the galactic emission and the scattered component.
As a result, we chose to use one of the two components based on whichever best improves the fit by minimizing the Cash statistic \citep[C;][]{Cash79}. 
In cases where neither the scattered nor galactic components improved the fit by a significant amount, both were excluded.
In order to be considered a significant improvement, the C statistic must decrease by more than the change in the degrees of freedom.

Once the model components were chosen for each source, parameter values were obtained using the Bayesian X-ray Analysis (BXA) package \citep{Buchner14_BXA}.
BXA utilizes
UltraNest\footnote{\url{https://johannesbuchner.github.io/UltraNest/}}---a Bayesian nested sampling package by \cite{Buchner21_ultranest}---to thoroughly search the parameter space.
This careful statistical treatment is necessary to avoid falling in local minima, which the standard \textsc{xspec} fitting routines are prone to do, especially in the complex parameter spaces created by physically realistic torus models.
BXA produces a posterior distribution which contains probability distribution functions (PDFs) for all free parameters.
Below, we describe the priors used for fitting the free parameters.

Where the baseline and clumpy models share the same free parameters, we adopt the same priors for both. 
$N_{\mathrm{H}}$ is assigned a log-uniform prior between $1.0 \times 10^{20}$~cm$^{-2}$ and $1.0 \times 10^{26}$~cm$^{-2}$.
$\Gamma$ is assigned a Gaussian prior centered at $1.8$---which is the average value for AGN \citep[e.g.][]{Nandra94,Ricci17-BAT}---with a standard deviation of $\lambda = 0.15$.
For the clumpy model, a log-uniform prior is used for the powerlaw normalization (min~$=10^{-10}$, max~$=0.1$).
For the baseline model, we instead keep the powerlaw normalization fixed and use a \textsc{cflux} component, which calculates the flux of the given component (in this case, the 2-10~keV flux of the transmitted powerlaw). 
This enables the unabsorbed flux and luminosity to be directly measured.
The \textsc{cflux} parameter is already in log space, so it is assigned a uniform prior between $-15$ and $-10$.
If a galactic or scattered component is included for a source, a log-uniform prior is assigned for $f_{\textsc{scatt}}$ (min~$=10^{-5}$, max~$=0.1$) or $n_{\textsc{gal}}$ (min~$=10^{-10}$, max~$=0.1$), respectively.

As a supplement to the BXA results, frequentist best-fit models were obtained using the standard routines in HEASOFT's \textsc{PyXSPEC}.
Fits are found using a semi-automated routine, which is described below.

The process for fitting the baseline model begins by freezing $N_{\mathrm{H}}$ to a negligible value, then constraining $\Gamma$ and the normalization of the transmitted powerlaw spectrum. 
$\Gamma$ is then frozen to the best-fit value, and constraints are found on $N_{\mathrm{H}}$.
Lastly, $\Gamma$ is freed in order to fit both $\Gamma$ and $N_{\mathrm{H}}$ to the spectrum.
For cases where $\Gamma$ is unconstrained or unrealistically low for an AGN ($\Gamma < 1.5$), it is fixed to 1.8 and constraints on $N_{\mathrm{H}}$ are obtained.

We employ a slightly different fitting routine for the clumpy model.
The method follows the suggestion of \cite{Buchner19_clumpy} by first fitting the clumpy model to the hard-band data (20-24~keV) with $\Gamma$, $N_{\mathrm{H}}$, and the normalization as free parameters.
This focuses on fitting the intrinsic powerlaw and avoids hitting deep local minima in the parameter space.
After obtaining initial constraints on $N_{\mathrm{H}}$ and $\Gamma$, the energy range is increased gradually (down to 15~keV, 8~keV, 5~keV, then all the data), and re-fit until the full NuSTAR (3-24~keV) and XMM-Newton (0.1-10~keV) bands are included.
If $\Gamma < 1.5$, this process is repeated with $\Gamma$ frozen to 1.8, then $\Gamma$ is freed in the final step.
In the case that $\Gamma$ is still unrealistically low, we retain the model with $\Gamma = 1.8$.

\begin{figure}
    \centering
    \includegraphics[width=0.45\textwidth]{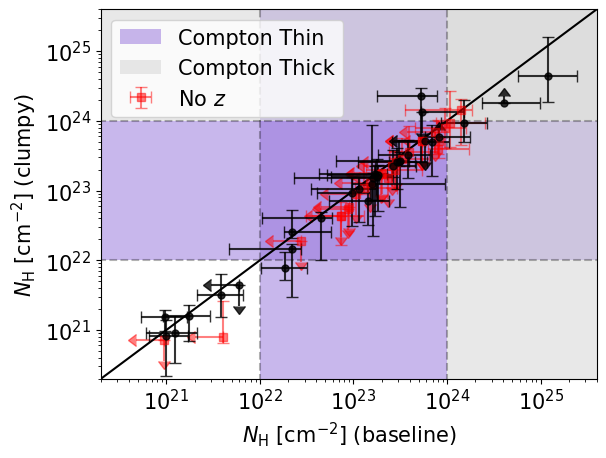}
    \caption{A comparison of the best-fit column densities ($N_{\mathrm{H}}$) for the baseline and clumpy models. 
    The shaded regions mark the Compton Thin (purple) and Compton Thick (gray) regimes.
    Sources without redshift measurements are shown as red squares.}
    \label{fig:nh_SimpVClumpy}
\end{figure}

For both the baseline and clumpy models, the final step of the automated routine is to fold in either the reflection component or the galactic emission, if either was used in the BXA fitting.
Manual inspection allowed for direct comparison between parameter values of the two models. 
Where discrepancies emerged, the source of the discrepancy was identified and the model was manually re-fit. 
At the end of this process, the baseline and clumpy models yield consistent $N_{\mathrm{H}}$ values, as demonstrated by Figure \ref{fig:nh_SimpVClumpy} (AGN 29, which has 
$N_{\mathrm{H}} \lesssim 10^{19}$ 
for both models, has been excluded from this plot for clarity).

More detail about individual fits---including the BXA results and frequentist best-fit parameters for each AGN---are shown in Appendix \ref{apendix-fits}.


\section{RESULTS}\label{sec:results}
\begin{figure}
    \centering
    \includegraphics[width=0.49\textwidth]{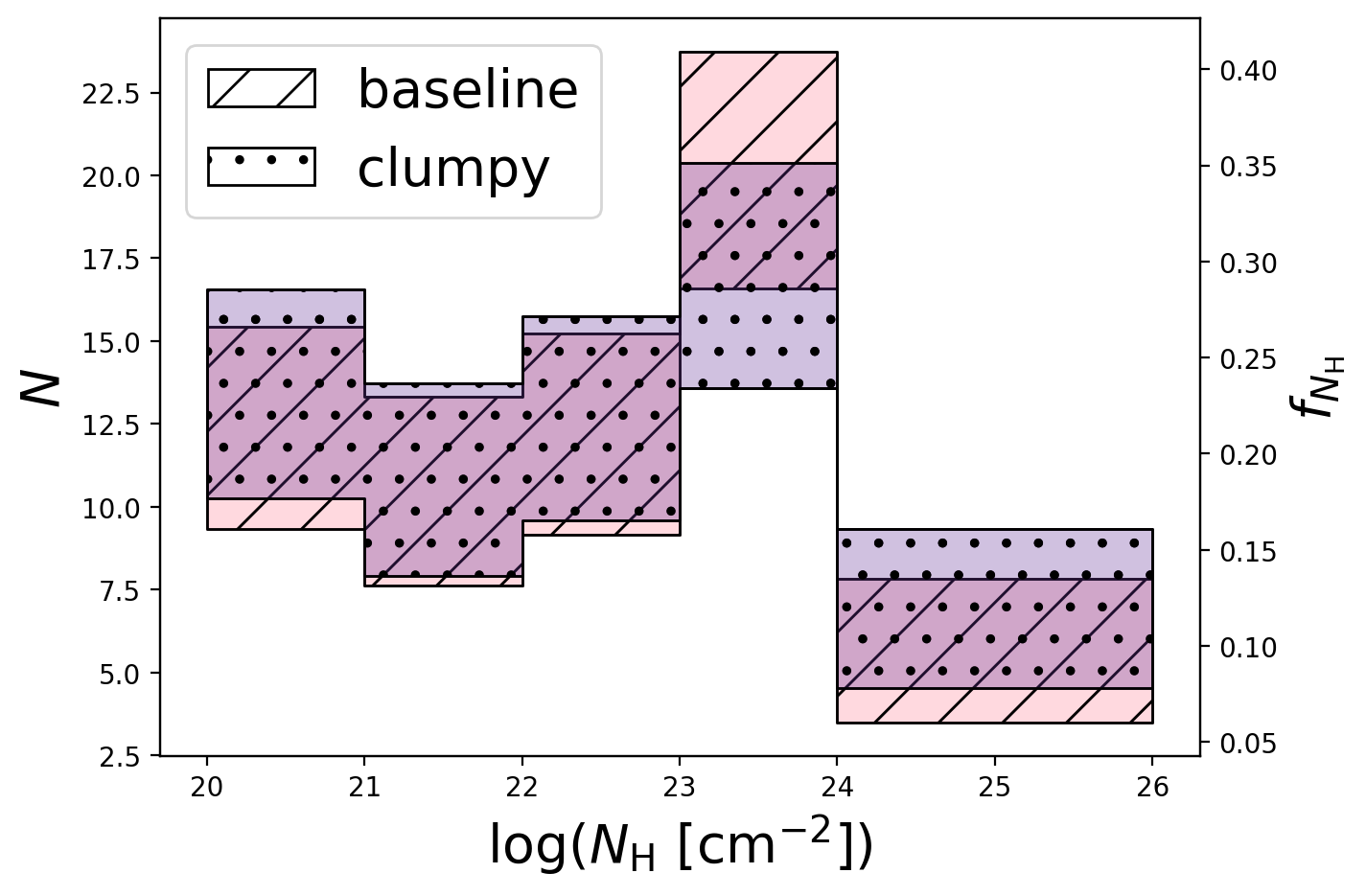}
    \caption{\mc{The effective number $N$ and fraction $f_{N_\mathrm{H}}$ of AGN observed in each log($N_{\mathrm{H}}$) bin within $1 \sigma$ for the baseline (pink; hatched) and clumpy (purple; dotted) models, as derived by integrating the summed PDF shown in Figure \ref{fig:nh_pdf} (Appendix \ref{apendix-fits}).}}
    \label{fig:nh_obs_bins}
\end{figure}

\begin{figure}
    \centering
    \includegraphics[width=0.49\textwidth]{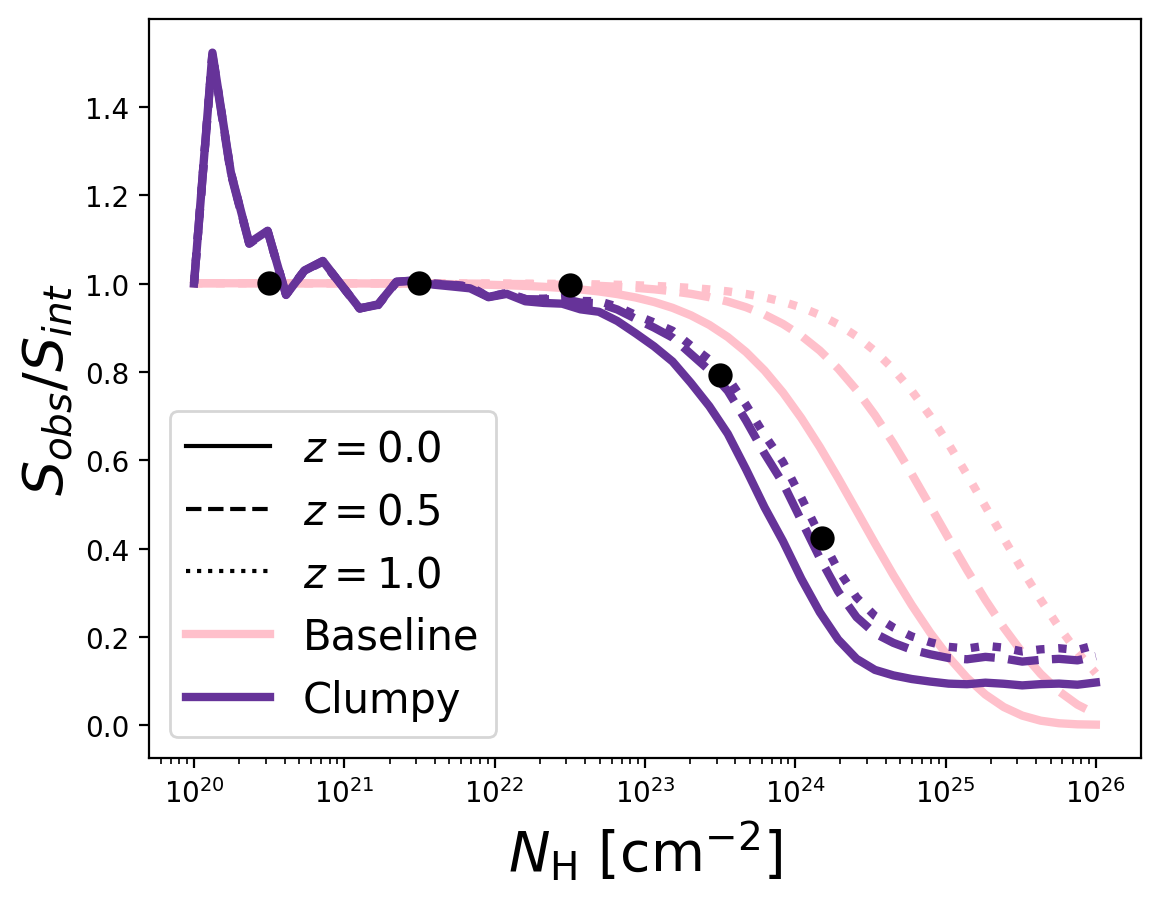} 
    \caption{The ratio of observed to absorption-corrected 8-24 keV flux ($S_{\text{obs}}/S_{\text{int}}$) as a function of absorption. The flux ratios are shown for the clumpy (purple) and baseline (pink) models at $z=0.0$ (solid lines), $0.5$ (dashed lines), and $1.0$ (dotted lines). The black points denote the ratios used in Equation \ref{eq:dNdH_exp} for each log($N_{\mathrm{H}}$) bin (see Table \ref{tab:logN_logS}).}
    \label{fig:obs/int}
\end{figure}

For the following discussion, two objects (ID 3 and 18) have been excluded. 
ID 3 was identified as a star by \citetalias{Zhao24-cycle5-6}, and a trustworthy fit could not be found for AGN 18 (as described in Appendix \ref{apendix-fits}).

\subsection{Observed $N_{\mathrm{H}}$ Distribution }\label{sec:nh_obs}

\begin{figure}
    \centering
    \includegraphics[width=0.4\textwidth]{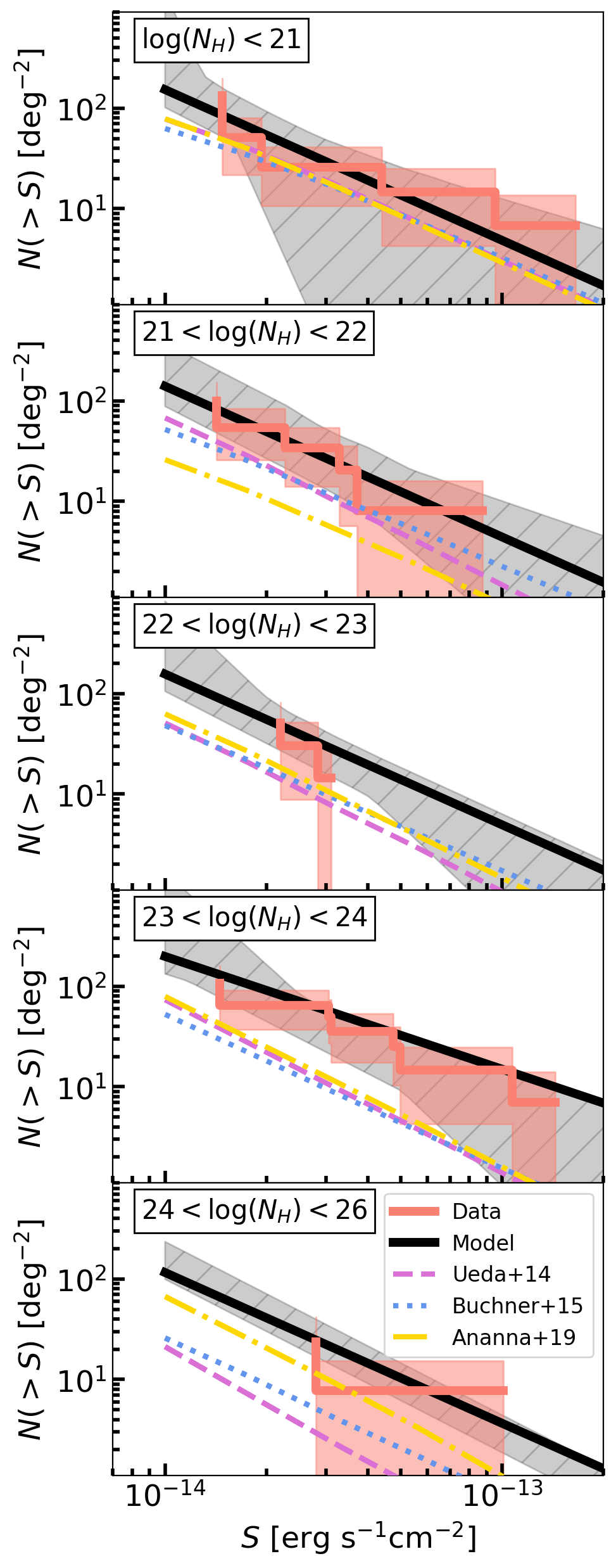}
    \caption{The median 8-24~keV logN-logS models (black lines) and $3\sigma$ confidence intervals (hatched regions) are shown in each log$(N_{\mathrm{H}})$ bin over 1000 iterations of the sample, as described in Section \ref{sec:nh_int}. \mc{The nominal data (binned using the median $N_{\mathrm{H}}$ from the posterior) are plotted in solid pink with poisson error bars (shaded region). logN-logS relations from population synthesis models \citep{Ueda14-CXB, Buchner15-PopSynth, Annana19-Accretion_HistoryI} are shown for comparison.}}
    \label{fig:logNlogS}
\end{figure}

In order to have a careful statistical treatment of our analysis, we chose to use the Bayesian results for calculating the $N_{\mathrm{H}}$ distribution of the sample. 
Following the methodology presented by \cite{Lanzuisi18-COSMOS_obscured}, the observed $N_{\mathrm{H}}$ distribution is found by extracting the posteriors produced by BXA (Section \ref{sec:BXA}). 
Smoothed PDFs were created for each source by applying a Gaussian Kernel Density Estimator (KDE; bandwith = 0.01) to $N_{\mathrm{H}}$ and normalizing so that the total integral of the curve is equal to 1.
Adding the PDFs from all sources then yields the observed $N_{\mathrm{H}}$ distribution of the sample. 
\mc{Figure \ref{fig:bxa_all} in Appendix \ref{apendix-fits} shows the individual PDFs of each source, and Figure \ref{fig:nh_pdf} shows the summed PDF of the sample.}

\mc{The sample is then divided into $1$~dex log$(N_{\mathrm{H}}~[$cm$^{-2}]/$cm$^{-2})$ (log$(N_{\mathrm{H}})$ for brevity) bins, with one bin representing the CT regime (log($N_{\mathrm{H}}) > 24$). 
To estimate the number $N$ and fraction $f_{N_\mathrm{H}}$ of AGN in each bin, the summed PDF is integrated within the bin.
1$\sigma$ confidence intervals are calculated using the process outlined by \cite{Cameron11_Baysean}, which uses the beta distribution to estimate confidence intervals on binomial population proportions \citep[see Appendix A of][]{Cameron11_Baysean}, where the number of successes is defined as the effective sample within the bin ($N$), and the sample size is the number of AGN used in the analysis (58).
This process gives estimates for observed $f_{\text{C-Thin}}$ (fraction of sources with $N_{\mathrm{H}} > 10^{22}~$cm$^{-2}$) and $f_{CT}$.}
For the baseline/clumpy model, this yields $f_{\text{C-Thin}}$ of \mc{$0.54^{+0.06}_{-0.06}/0.49^{+0.06}_{-0.06}$} and \mc{$f_{\text{CT}} = 0.09^{+0.02}_{-0.05}/0.11^{+0.03}_{-0.05}$}.


\subsection{Intrinsic $N_{\mathrm{H}}$ Distribution }\label{sec:nh_int}

Even with NuSTAR's hard energy range, intrinsically faint AGN become more difficult to detect in the CT regime due to emission being heavily attenuated \citep[e.g.][]{Burlon11-BAT}; as shown in Figure \ref{fig:obs/int}, much of the X-ray emission in the 8-24 keV band is suppressed at high absorption.
In order to account for this absorption bias, we follow the process outlined by \cite{Burlon11-BAT} and \cite{Zappacosta18-NUSTAR_extragal_hard-band} and derive an estimate of the true, intrinsic $N_{\mathrm{H}}$ ($dN/d\text{log}N_{\mathrm{H}}$) distribution by binning the sample into log($N_{\mathrm{H}}$) bins and integrating the log$N$-log$S$ relation ($dN/dS$) of each bin over an appropriate flux range:

\begin{equation}
    \frac{dN}{d\text{log}N_{\mathrm{H}}} = 
    \int_{S_{\text{min}}^{\text{obs}}}^{S_{\text{max}}^{\text{obs}}}
    \frac{dN}{dS}~dS.
\label{eq:dNdH}
\end{equation}

This analysis is performed in the $8 - 24$~keV band.
Only sources detected above the $95\%$ reliability threshold and with signal to noise ratios $>2.5$ are used in order to ensure accurate log$N$-log$S$ modeling.
This reduces the sample to $24$ objects (see Table \ref{tab:pars} in Appendix \ref{apendix-fits} and Table 2 in \citetalias{Zhao24-cycle5-6}).
In the following steps, we describe our procedure for acquiring the parameters in Equation \ref{eq:dNdH}.

The first step is to select 1000 random realizations from the posteriors produced by BXA.
The clumpy model results are only used for objects whose clumpy and baseline posteriors are both $\geq 90 \%$ above log$(N_{\mathrm{H}}) = 23$. 
Beyond this threshold, the physically motivated components of obscured AGN become significant \citep{Yaqoob12_MYTorus, Balokovic18_borus, Buchner19_clumpy}.
The clumpy model is unnecessarily complex for less obscured spectra, so all other sources are assigned their baseline posteriors.

For each realization, we divide the sample into $1$~dex log($N_{\mathrm{H}}$) bins, create cumulative flux distributions (log$N$-log$S$) per bin, and fit the log$N$-log$S$ curves to powerlaw functions ($N(>S) = A(N_{\mathrm{H}}) S^{-\alpha}$, where $\alpha$ is the powerlaw slope and $A(N_{\mathrm{H}})$ is the normalization at $10^{13}$ erg/s/cm$^2$).
The slope for a Euclidean universe ($\alpha = 3/2$) is assumed for bins with $< 5$ sources.

Each log$(N_{\mathrm{H}})$ bin is assigned the median $A(N_{\mathrm{H}})$ and $\alpha$ from the 1000 realizations.
In Figure \ref{fig:logNlogS}, these median logN-logS models are plotted alongside the data (which are assigned the median $N_{\mathrm{H}}$ from their posterior distribution) and compared to \citet{Ueda14-CXB}, \citet{Buchner15-PopSynth}, and \citet{Annana19-Accretion_HistoryI}.
In each log($N_{\mathrm{H}}$) bin, we find $\gtrsim 2$ times more sources than are predicted by these population synthesis models, which is consistent with the logN-logS curve of the 8-24~keV sample as a whole (see the middle panel of Figure 20 in \citetalias{Zhao24-cycle5-6}).

Before performing the integral in Equation \ref{eq:dNdH}, the flux range must be carefully selected.
In particular, minimum observed fluxes ($S_{\text{min}}^{\text{obs}}$) must be chosen
so that the intrinsic (absorption-corrected) flux $S_{\text{min}}^{\text{int}}$ is the same in each log($N_{\mathrm{H}}$) bin.
That way, we can be relatively confident that all AGN down to $S_{\text{min}}^{\text{int}}$ are being detected regardless of $N_{\mathrm{H}}$.

To select $S_{\text{min}}^{\text{int}}$, we start by finding the minimum observed flux that appears in the CT bin (log$(N_{\mathrm{H}})> 24$).
The CT bin for this sample is sensitive down to $S_{\text{obs}} \approx 3.0 \times 10^{-14}$~erg/s/cm$^{2}$, which corresponds to an unabsorbed flux of $S_{\text{int}} \approx 6.0 \times 10^{-14}$~erg/s/cm$^{2}$, 
so this becomes $S_{\text{min}}^{\text{int}}$.
In each $N_{\mathrm{H}}$ bin, this is converted to an observed flux $S_{\text{min}}^{\text{obs}} = k(N_{\mathrm{H}}) S_{\text{min}}^{\text{int}}$, where $k(N_{\mathrm{H}})$ is pulled from $S_{\text{obs}}/S_{\text{int}}$ curves shown in Figure \ref{fig:obs/int}.
For our chosen value of $\Theta_{\text{inc}}$, the $k(N_{\mathrm{H}})$ of \textsc{uxclumpy} behaves erratically at log$(N_{\mathrm{H}}) < 23 $ (Figure \ref{fig:obs/int}). 
Since we only use the clumpy model for log$(N_{\mathrm{H}}) > 23 $, this does not affect our analysis.

In order to compute $k(N_{\mathrm{H}})$, log$(N_{\mathrm{H}})$ is set to the middle of each bin.
However, the CT bin has a large range of $k(N_{\mathrm{H}})$, so we chose to calculate $k(N_{\mathrm{H}})$ at $N_{\mathrm{H}} = 1.5 \times 10^{24}$~cm$^{-2}$, which is the formal definition of the CT threshold and gives a conservative estimate of $f_{CT}$
\citep[as done by][]{Zappacosta18-NUSTAR_extragal_hard-band}. A scattered component with $f_{scatt} = 10.9 \times N_{\mathrm{H}}^{-0.47}$ is assumed \citep[see Table 1 from][]{Gupta21},
and the redshift is set to the median redshift of objects in that bin.

Table \ref{tab:logN_logS} reports the median redshifts, $k(N_{\mathrm{H}})$, and logN-logS parameter values.
These parameters go into Equation \ref{eq:dNdH} after performing the integral:

\begin{equation}
    \frac{dN}{d\text{log}N_{\mathrm{H}}} = 
    \frac{A(N_{\mathrm{H}})}{(10^{-13})^{-\alpha}}
    \biggl[
    (S_{\text{max}}^{{\text{obs}}})^{-\alpha} - 
    (S_{\text{min}}^{{\text{obs}}} k(N_{\mathrm{H}}))^{-\alpha}
    \bigg].
\label{eq:dNdH_exp}
\end{equation}

\begin{table}[t]
    \centering
    \caption{Best-fit logN-logS parameters.\label{tab:logN_logS} The normalization is defined as the value at $10^{-13}$ $(A(N_{\mathrm{H}}) = N( >S(3-24$~keV)$ = 10^{-13}$~erg/s/cm$^2))$}
    \begin{tabular}{ccccc}
     \hline
     \hline 
     \\[-0.75em]
      log($N_{\mathrm{H}}$) & $<z>$ & $\alpha$ & $A(N_{\mathrm{H}})$ & $k(N_{\mathrm{H}})$ \\\\[-0.75em]
     	\hline
	\hline
	20-21 & 0.65 & $1.50_{-0.6}^{+5.0}$ & $5.2_{-5}^{+11}$ & 1.00 \\
    21-22 & 0.58 & $1.50_{-0.5}^{+0.9}$ & $4.5_{-3}^{+10}$ & 1.00 \\
    22-23 & 0.57 & $1.50_{-0.2}^{+1.3}$ & $5.1_{-4}^{+5}$ & 1.00 \\
    23-24 & 0.62 & $1.12_{-0.3}^{+1.5}$ & $12.5_{-10}^{+8}$ & 0.79 \\
    24-26 & 0.86 & $1.50_{-0.0}^{+0.0}$ & $3.7_{-1}^{+4}$ & 0.42 \\
 
      \hline
    \end{tabular}
\end{table}

\begin{figure*}
 \centering
\includegraphics[width=0.8\textwidth]{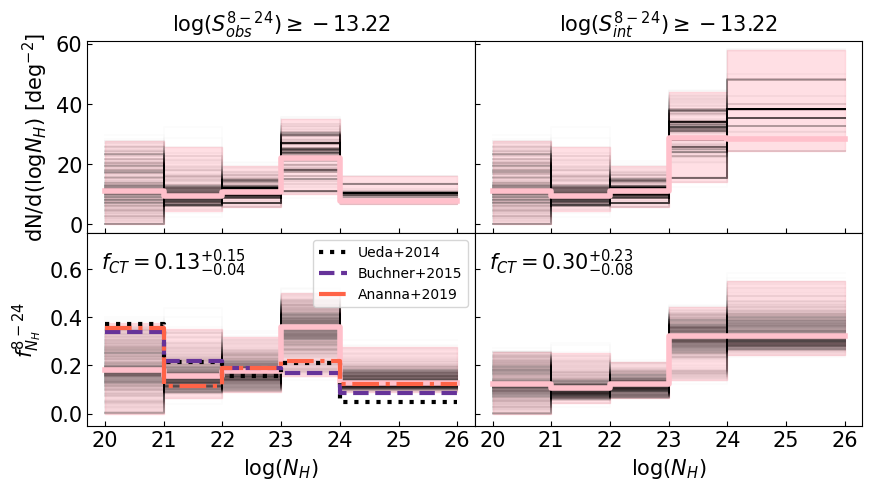}
    \caption{The $N_{\mathrm{H}}$ distribution (top) and $N_{\mathrm{H}}$ fraction (bottom) of the sample down to an observed (left) and absorption-corrected (right) 8-24~keV flux of $6.0 \times 10^{-14}$~erg/s/cm$^{2}$. The distribution has been calculated over 1000 realizations, all plotted in black, and the median and $3 \sigma$  errors of the distribution are plotted in pink. The predicted observed $f_{N_{\mathrm{H}}}$ for our sample is shown for three population synthesis models \citep{Ueda14-CXB, Buchner15-PopSynth, Annana19-Accretion_HistoryI} in the bottom-left figure.} 
    \label{fig:dNdH}
\end{figure*}

Figure \ref{fig:dNdH} shows the resulting measurement of the intrinsic $N_{\mathrm{H}}$ distribution alongside all 1000 realizations of the sample.
Correcting for absorption bias and using a minimum unabsorbed flux of $S^{int}_{min} = 6.0 \times 10^{-14}$~erg/s/cm$^{2}$ yields $f_{CT} = 0.30_{-0.08}^{+0.23}$.
For comparison, the $N_{\mathrm{H}}$ distribution is also shown without accounting for absorption bias, but using the same flux limit. 
This gives $f_{CT} = 0.13_{-0.04}^{+0.15}$.


\section{DISCUSSION}\label{sec:discussion}
\begin{figure}
 \centering
    \includegraphics[width=0.45\textwidth]{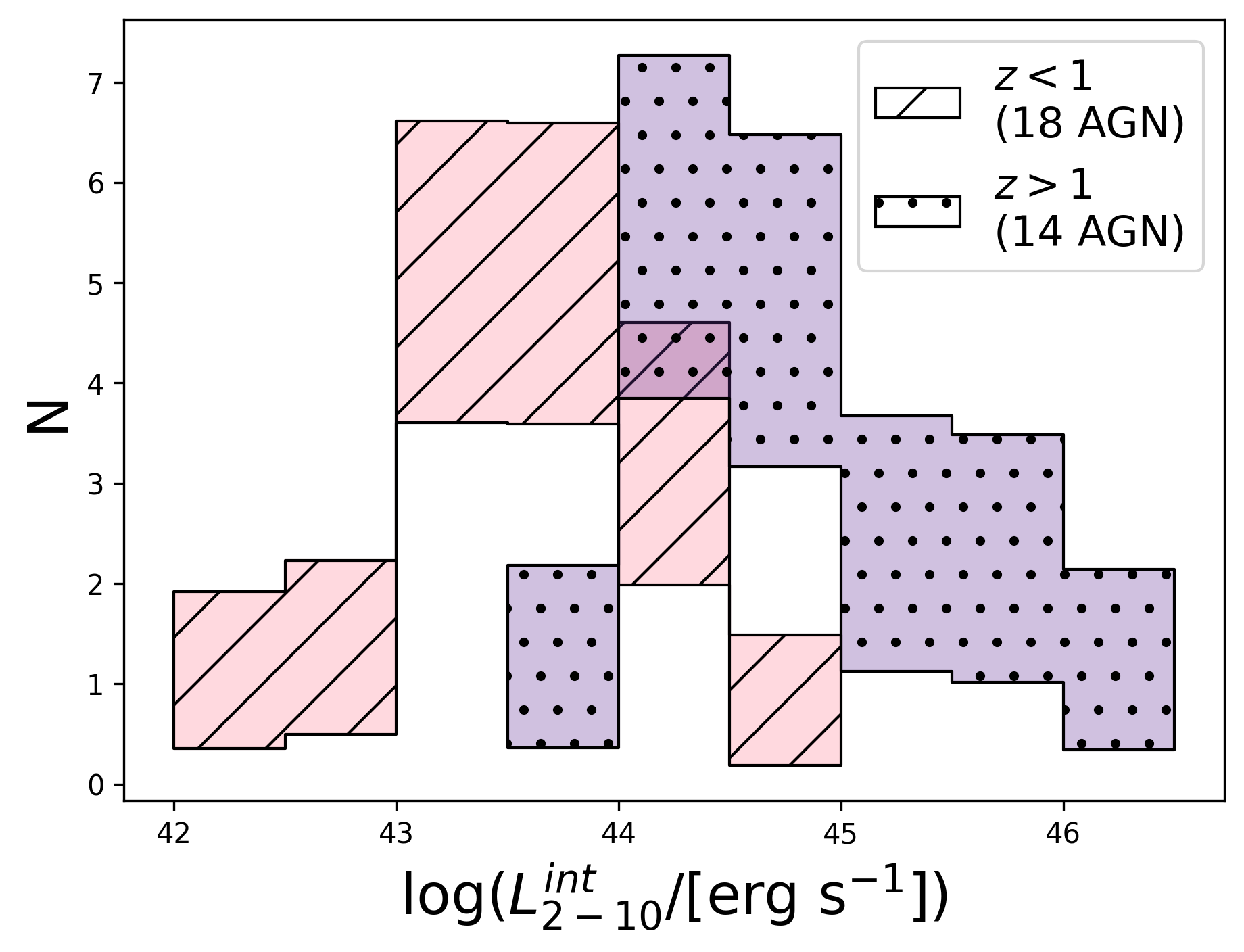}
    \caption{\mc{The unabsorbed 2-10~keV luminosity distribution of sources with redshift measurements. The sample is split into low ($z<1$; pink, hatched) and high ($z>1$; purple, dotted) redshift bins. The shaded region gives $1 \sigma$ confidence intervals.}}
    \label{fig:L_int}
\end{figure}

\begin{figure}
 \centering
    \includegraphics[width=0.45\textwidth]{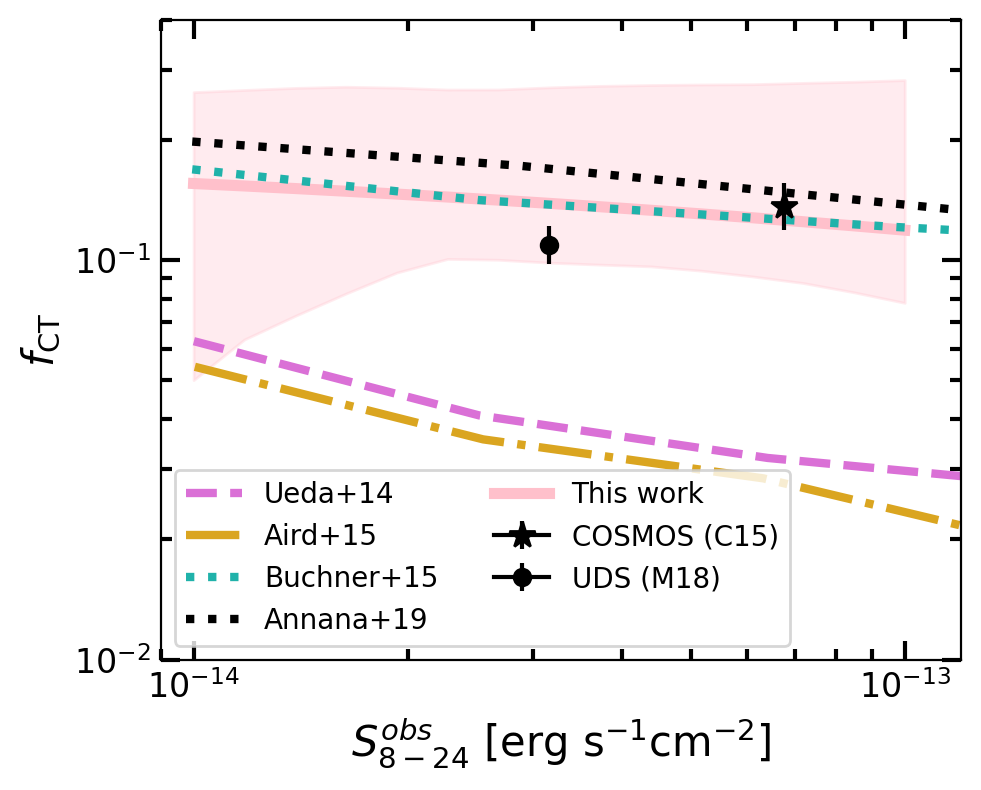}
    \caption{$f_{CT}$ as a function of observed 8-24~keV flux (solid pink), with the shaded region denoting $3 \sigma$ error bars for this work. For comparison, predictions from population synthesis models \citep[$z < 3$;][]{Ueda14-CXB, Aird15,Buchner15-PopSynth, Annana19-Accretion_HistoryI} and $f_{CT}$ measurements from the NuSTAR COSMOS \citep[black star]{Civano15-NuSTAR_COSMOS} and UDS \citep[black dot]{Masini18-NuSTAR_UDF} surveys are shown.}
    \label{fig:f_v_S}
\end{figure}

\begin{figure}
 \centering
    \includegraphics[width=0.45\textwidth]{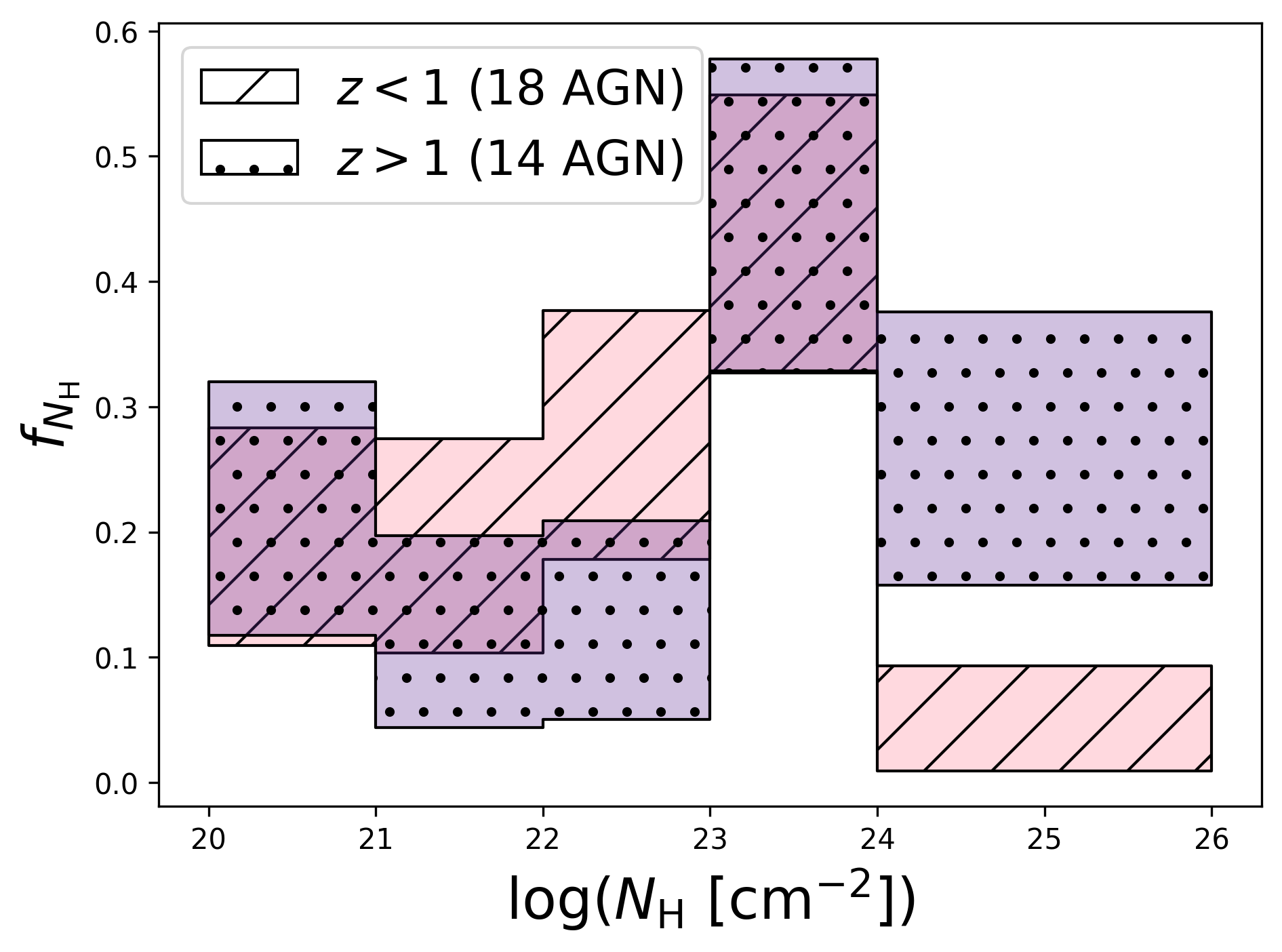}
    \caption{The observed column density distribution of the sample for low ($z < 1$; pink, hashed) and high ($z > 1$; purple, dotted) redshifts.}
    \label{fig:nh_pdf_z}
\end{figure}

\subsection{Comparison to previous measurements}

In every column density bin, we find more sources than are expected from population synthesis models \citep[][see Figure \ref{fig:logNlogS}]{Ueda14-CXB,Buchner15-PopSynth, Annana19-Accretion_HistoryI}. However, our observed $f_{CT}$ is consistent with \citet{Annana19-Accretion_HistoryI} (Figure \ref{fig:dNdH}).
Table \ref{tab:f_CT} compares our absorption-corrected calculation of $f_{CT}$ to measurements found in previous surveys  \citep{Wilkes13, Civano15-NuSTAR_COSMOS, Lanzuisi18-COSMOS_obscured, Masini18-NuSTAR_UDF, Zappacosta18-NUSTAR_extragal_hard-band, Kura21, TA21-CT, Carroll23, Akylas24-CT_local, Boorman25}, and the latest population synthesis models \citep{Annana19-Accretion_HistoryI, Gerolymatou25}. In all cases, the measurements reported in Table \ref{tab:f_CT} use $f_{CT} = N_{CT}/N_\text{total}$, where \citet{Wilkes13} and \citet{Kura21} use $1.5 \times 10^{24}$~cm$^{2}$ as their CT threshold and all other studies (including this work) use $10^{24}$~cm$^{2}$.
Our finding is consistent with these previous studies, though with large error-bars.
The luminosity distributions of these studies are also listed. 
For this work, the intrinsic (unabsorbed) 2-10~keV luminosity distribution of sources with redshift measurements is shown in Figure \ref{fig:L_int}.

Future work could improve this measurement by including the full Cycle 5+6+8+9 catalog \citep{Silver25} and XMM-Newton identified  sources.
Additionally, the measured value of $f_{CT}$ can vary dramatically based on the model used \citep{Boorman25}, and incorporating a larger variety of models would yield a more robust measurement of $f_{CT}$.

Some unusual features appear in the log($N_{\mathrm{H}}$) distribution. 
In particular, we find a small fraction ($\sim 20\%$) of AGN in the log$(N_{\mathrm{H}}$/cm$^{-2}) = 20-21$ bin, which conflicts with most of the studies shown in Table \ref{tab:f_CT} and Figure \ref{fig:dNdH}.
Additionally, the distribution is relatively flat at  log\mc{$(N_{\mathrm{H}}) < 23$}.
This may be due to the lack of XMM-Newton matches for many objects.
Without XMM-Newton, the degree of obscuration is difficult to constrain at low values, leading to flat PDFs (see Figure \ref{fig:bxa_all} in Appendix \ref{apendix-fits}).

In order to derive the relationship between $f_{CT}$ and the observed 8-24~keV flux, $f_{CT}$ and $3\sigma$ confidence intervals were calculated for different values of $S_{8-24}^{obs}$ using the logN-logS models in Figure \ref{fig:logNlogS}.
Figure \ref{fig:f_v_S} shows the results compared to population synthesis models up to $z>3$ \citep{Ueda14-CXB, Aird15, Buchner15-PopSynth, Annana19-Accretion_HistoryI} and previous measurements from the NuSTAR COSMOS \citep{Civano15-NuSTAR_COSMOS} and UDS \citep{Masini18-NuSTAR_UDF} surveys. 
Within uncertainties, the data are consistent with \citet{Buchner15-PopSynth} and \citet{Annana19-Accretion_HistoryI}.

\tabletypesize{\large}
\begin{table*}
\begin{center}
        
    \caption{Measurements  of  $f_{CT}$ \label{tab:f_CT}}
    \begin{tabular}{llll}
     \hline 
     \hline
     \\[-0.75em]
     \multicolumn{1}{c}{Work} & \multicolumn{1}{c}{$z$} & \multicolumn{1}{c}{$L_X$$^a$} & \multicolumn{1}{c}{$f_{CT}$}\\ 
     	\hline
	\hline
	Ananna+ 19 & $\leq 1$ & $10^{41} < L_{2-10}^{int} < 10^{47}$ & $0.56 \pm 0.07$ \\ 
    
	Gerolymatou+ 25 & $\leq 3$ & $10^{41.5} < L_{2-10}^{int} < 10^{46.5}$ & $0.21 \pm 0.07$ \\ 
	
     \hline
     Wilkes+ 13  & $1.0-2.0$ & $10^{43} < L_{2-8}^{int} < 10^{45.75}$ & $0.21 \pm 0.07$ \\
     Civano+ 15 & $\sim 0.5$ & $10^{42} < L_{10-40}^{obs} < 10^{45.5}$ & $0.13 \pm 0.03$ \\
     Lanzuisi+ 18 & $0.1-1$ & $10^{41.5} < L_{2-10}^{int} < 10^{45.5}$ & $\sim 0.2 $ \\
                          & $1-2$ & & $\sim 0.3 $ \\
                          & $2-3$ & & $\sim 0.5 $ \\
     Zappacosta+ 18 & $\sim 0.5$ & $10^{42.6} < L_{10-40}^{int} < 10^{45.6}$ & 0.02-0.56 \\ 
     Masini+ 18 &  $\sim 1$ & $10^{42.1} < L_{10-40}^{obs} < 10^{46}$ & $0.115 \pm 0.020$ \\ 
     Kuraszkiewicz+ 21 & $0.5-1.0$ & $10^{41} < L_{2-10}^{int} < 10^{47}$ & $\sim 0.2$ \\
     Torres-Albà+ 21 & $\leq 0.05$ & $10^{42} \lesssim L_{2-10}^{int}\lesssim 10^{44}$ & $\sim0.08$ \\ 
     Carroll+ 23 & $\leq 0.8$ & $10^{42.5} < L_{2-10}^{int} < 10^{44.7}$ & $0.555^{+0.037}_{-0.032}$ \\
     Akylas+ 24 & $< 0.02$  & $10^{39.5} < L_{2-10}^{int} < 10^{44.5}$$^{(b)}$ & $0.25 \pm 0.05$ \\
     Boorman+ 25 & $\leq 0.044$ & $10^{41} \lesssim L_{2-10}^{int}\lesssim 10^{44} $ & $0.35 \pm 0.09$\\ 
     \hline
     This work & $\sim 0.5$ & $10^{42} \lesssim L_{2-10}^{int}\lesssim 10^{46}$ & $0.30_{-0.08}^{+0.21}$ \\
 
      \hline
    \end{tabular}
    \end{center}
    
$^a$ X-ray Luminosity distribution reported by the study in units of erg~s$^{-1}$. Superscripts indicate the energy band (in keV) and subscripts define whether the reported luminosities are observed or intrinsic (corrected for absorption).

$^b$ Only the luminosities of sources detected in the X-ray are reported.
\end{table*}

\subsection{Redshift evolution}

\mc{As discussed in Section \ref{sec:samp}, 33 of the sources in this work have multi-wavelength matches with redshift measurements.
While this sample is incomplete, there are enough sources to perform a preliminary test of previous studies \citep{Lanzuisi18-COSMOS_obscured,Peca23}, which find that obscured AGN dominate at higher redshifts.}

\mc{After splitting the sample into low ($z < 1$) and high ($z > 1$) redshift bins, we use the same method described in Section \ref{sec:nh_obs}---using the baseline model for sources with $N_{\mathrm{H}} < 23$ and the clumpy model for $N_{\mathrm{H}} \geq 23$---to derive the observed $N_{\mathrm{H}}$ distribution shown in Figure \ref{fig:nh_pdf_z}.
A two sample KS test is used to assess whether the low and high redshift samples are significantly different.
Iterating over 1000 realizations of the sample drawn from the posteriors (as described in Section \ref{sec:nh_int}) and performing the KS test in each realization yields an average p-value of 0.023.
This p-value suggests that the samples are statistically different, with more CT objects falling in the $z > 1$ bin.
We observe $f_{CT} < 0.09$ for the low redshift sample and $f_{CT} = 0.23^{+0.14}_{-0.08}$ for the $z > 1$ sample.
However, this does not account for luminosity and redshift-dependent biases.
Accounting for these biases is non-trivial \citep[i.e, see][]{Peca23}, and we leave it to future work as spectroscopic campaigns yield a more complete sample of redshifts.}

\section{SUMMARY AND CONCLUSIONS}\label{sec:conclusion}
We derive the $N_{\mathrm{H}}$ distribution of the hard X-ray detected sources identified in the \citetalias{Zhao24-cycle5-6} NuSTAR-NEP catalog. We use two models---a baseline absorbed powerlaw model and a clumpy torus model \citep{Buchner19_clumpy}--- and we employ a Bayesian treatment \citep{Buchner14_BXA} to fully encapsulate the complicated parameter spaces of our models.  We summarize our findings below.

\begin{itemize}

    \item We measure a Compton Thick fraction of \mc{$0.13_{-0.04}^{+0.15}$/$0.30_{-0.08}^{+0.21}$} down to an observed/unabsorbed flux of 
    $6.0 \times 10^{-14}$~erg/s/cm$^{2}$
    (see Figure \ref{fig:dNdH} and Table \ref{tab:f_CT}).
 
    \item Our Compton Thick fraction is consistent with population synthesis models and previous studies (see Table \ref{tab:f_CT} and Figure \ref{fig:dNdH}), though we find a flatter $N_{\mathrm{H}}$ distribution for unabsorbed sources. This may be explained by poor statistics, especially in the soft band for sources without XMM-Newton counterparts. Additionally, we are likely underestimating the $N_{\mathrm{H}}$ of the 27 sources that lack redshift measurements. 
    
    \item The measured relationship between $f_{CT}$ and the observed 8-24~keV flux is consistent with previous findings \citep{Buchner15-PopSynth,Civano15-NuSTAR_COSMOS,Masini18-NuSTAR_UDF, Annana19-Accretion_HistoryI}, though with large uncertainties.

    \item From the 33 sources with redshift measurements, we see evidence that the $N_{\mathrm{H}}$ distribution may evolve with redshift, with an observed CT fraction $f_{CT} < 0.09$ for low redshift sources ($z < 1$) and $f_{CT} = 0.23^{+0.14}_{-0.08}$ for high-redshift sources ($z > 1$). 
    More complete redshift measurements and a thorough account of redshift and luminosity-dependent biases are needed to investigate this finding.
    
\end{itemize}


The NEP field has some of the deepest X-ray data available, and it is the only deep survey designed to have simultaneous soft and hard X-ray observations. However, this type of simultaneous program is resource-intensive, and observational constraints on faint AGN in the NEP field come with large uncertainties. This demonstrates that a new telescope with high angular resolution and broad-band X-ray spectroscopy \citep[such as the previously proposed X-ray probe-class mission HEX-P;][]{Boorman24-HEXP,Civano24-HEXP,Garcia24-HEXP,Kammoun24-HEXP,Pfeifle24-HEXP,Piotrowska24-HEXP} would be an invaluable asset to future studies of AGN in the X-ray band.

\section*{Acknowledgments}
The material is based upon work supported by NASA under award numbers 80GSFC21M0002 and 80NSSC24K1482.
This work made use of data from the NuSTAR mission, which is led by the California Institute of Technology, managed by the Jet Propulsion Laboratory, and funded by the National Aeronautics and Space Administration.
Additionally, this work made use of data from XMM-Newton, an ESA science mission with instruments and contributions directly funded by ESA Member States and NASA.
CNAW acknowledges funding from  JWST/NIRCam contract to the University of Arizona NAS5-02015.

\mc{This research used data obtained with the Dark Energy Spectroscopic Instrument (DESI). DESI construction and operations is managed by the Lawrence Berkeley National Laboratory. This material is based upon work supported by the U.S. Department of Energy, Office of Science, Office of High-Energy Physics, under Contract No. DE–AC02–05CH11231, and by the National Energy Research Scientific Computing Center, a DOE Office of Science User Facility under the same contract. Additional support for DESI was provided by the U.S. National Science Foundation (NSF), Division of Astronomical Sciences under Contract No. AST-0950945 to the NSF’s National Optical-Infrared Astronomy Research Laboratory; the Science and Technology Facilities Council of the United Kingdom; the Gordon and Betty Moore Foundation; the Heising-Simons Foundation; the French Alternative Energies and Atomic Energy Commission (CEA); the National Council of Humanities, Science and Technology of Mexico (CONAHCYT); the Ministry of Science and Innovation of Spain (MICINN), and by the DESI Member Institutions: www.desi.lbl.gov/collaborating-institutions. The DESI collaboration is honored to be permitted to conduct scientific research on I’oligam Du’ag (Kitt Peak), a mountain with particular significance to the Tohono O’odham Nation. Any opinions, findings, and conclusions or recommendations expressed in this material are those of the author(s) and do not necessarily reflect the views of the U.S. National Science Foundation, the U.S. Department of Energy, or any of the listed funding agencies.}

This work also made extensive use of Astropy \citep{astropy13,astropy18,astropy22}, numpy \citep{numpy}, and SciPy \citep{SciPy}.
We would also like to thank the High Energy Astrophysics Science Archive Research Center (HEASARC) team; this work would not have been possible without their data and software.

We thank the anonymous referee for their comments, which greatly improved this paper. Lastly, we thank thank Steven Willner for his comments and and Luca Zappacosta for helpful discussion.

\bibliography{NEP_bib}{}
\bibliographystyle{aasjournal}

\newpage
\appendix
\section{NuSTAR source extraction regions}\label{sec:a_r}

\tabletypesize{\footnotesize}
\begin{deluxetable*}{c|cccc|cccc||c|cccc|cccc}\label{tab:snr_size}
\tablecaption{NuSTAR Spectral Extraction Region Sizes}
\startdata
 \\
  \multicolumn{2}{c}{}  &
 \multicolumn{2}{c}{FPMA} &
 \multicolumn{2}{c}{} & 
 \multicolumn{2}{c}{FPMB}  & &
  \multicolumn{2}{c}{} & 
 \multicolumn{2}{c}{FPMA} &
 \multicolumn{2}{c}{} & 
 \multicolumn{2}{c}{FPMB}  & \\
  \hline
  & &
  \multicolumn{2}{c}{Cycle} & & &
 \multicolumn{2}{c}{Cycle} & &
 & & 
  \multicolumn{2}{c}{Cycle} & & &
 \multicolumn{2}{c}{Cycle} &
 \\
  ID & 5 & 6 & 8 & 9 & 5 & 6 & 8 & 9 &
  ID & 5 & 6 & 8 & 9 & 5 & 6 & 8 & 9
  \\
  \hline
 1 & 20'' & - & 25'' & 15'' & 35'' & - & 15'' & 35'' &31 & 15'' & 20'' & 40'' & 10'' & 15'' & 20'' & 30'' & 15''\\
2 & 20'' & - & 35'' & 50'' & 30'' & - & 40'' & 30'' &32 & 15'' & 35'' & 45'' & 40'' & 20'' & 20'' & 45'' & 35''\\
3 & 35'' & 15'' & 15'' & 20'' & 20'' & 20'' & 15'' & 45'' &33 & 20'' & 20'' & 15'' & 15'' & 30'' & 20'' & 15'' & 20''\\
4 & 55'' & 25'' & 30'' & 45'' & 10'' & 50'' & 55'' & 40'' &34 & 20'' & 15'' & 15'' & 15'' & 25'' & 15'' & 35'' & 10''\\
5 & 20'' & 30'' & 50'' & 25'' & 20'' & 50'' & 35'' & 40'' &35 & 15'' & 10'' & 15'' & 30'' & 15'' & 10'' & 10'' & 30''\\
6 & 25'' & 25'' & 35'' & 30'' & 25'' & 25'' & 30'' & 25'' &36 & 40'' & 40'' & 45'' & 35'' & 25'' & 25'' & 40'' & 20''\\
7 & 25'' & 40'' & 30'' & 40'' & 20'' & 30'' & 25'' & 25'' &37 & 20'' & 15'' & 35'' & 40'' & 20'' & 25'' & 35'' & 40''\\
8 & 20'' & 30'' & 20'' & 10'' & 20'' & 10'' & 20'' & 30'' &38 & 60'' & 40'' & 40'' & - & 40'' & 45'' & 40'' & -\\
9 & 10'' & 20'' & 40'' & 10'' & 25'' & 25'' & 15'' & 10'' &39 & 20'' & 20'' & 30'' & 35'' & 30'' & 10'' & 30'' & 30''\\
10 & 15'' & 20'' & 20'' & 20'' & 20'' & 35'' & 30'' & 55'' &40 & 15'' & 25'' & 10'' & 20'' & 10'' & 20'' & 25'' & 25''\\
11 & 20'' & 20'' & 25'' & 25'' & 35'' & 20'' & 25'' & 35'' &41 & 40'' & 25'' & 25'' & 30'' & 35'' & 20'' & 30'' & 25''\\
12 & 25'' & 25'' & 15'' & 45'' & 35'' & 20'' & 25'' & 25'' &42 & 20'' & 15'' & 35'' & 30'' & 40'' & 10'' & 20'' & 55''\\
13 & 25'' & 25'' & 15'' & 20'' & 15'' & 20'' & 20'' & 30'' &43 & 25'' & 20'' & 45'' & 55'' & 15'' & 25'' & 35'' & 20''\\
14 & 25'' & 30'' & 40'' & 40'' & 15'' & 25'' & 40'' & 50'' &44 & 20'' & 15'' & 40'' & 45'' & 10'' & 30'' & 15'' & 15''\\
15 & 20'' & 30'' & 20'' & 10'' & 20'' & 20'' & 15'' & 10'' &45 & 20'' & 20'' & 20'' & 15'' & 20'' & 20'' & 25'' & 25''\\
16 & 15'' & 25'' & 40'' & 35'' & 55'' & 55'' & 30'' & 65'' &46 & 10'' & 20'' & 20'' & 20'' & 20'' & 20'' & 20'' & 15''\\
17 & 30'' & 15'' & 15'' & 20'' & 20'' & 40'' & 10'' & 20'' &47 & 20'' & 25'' & 35'' & 15'' & 20'' & 30'' & 45'' & 20''\\
18 & 10'' & 20'' & 25'' & 20'' & 40'' & 15'' & 50'' & 55'' &48 & 35'' & 20'' & 30'' & 35'' & 30'' & 25'' & 40'' & 25''\\
19 & 20'' & 25'' & 50'' & 25'' & 40'' & 45'' & 50'' & 20'' &49 & 15'' & 15'' & 25'' & 45'' & 15'' & 20'' & 10'' & 20''\\
20 & 25'' & - & 35'' & 45'' & 10'' & - & 10'' & 40'' &50 & 20'' & 20'' & 10'' & 35'' & 20'' & 30'' & 15'' & 10''\\
21 & 20'' & 20'' & 25'' & 30'' & 20'' & 20'' & 35'' & 20'' &51 & 20'' & 20'' & 20'' & 40'' & 20'' & 20'' & 20'' & 20''\\
22 & 20'' & 20'' & 10'' & 25'' & 20'' & 20'' & 10'' & 25'' &52 & 20'' & 30'' & 10'' & 20'' & 20'' & 15'' & 10'' & 15''\\
23 & 40'' & 10'' & 10'' & 10'' & 10'' & 40'' & 40'' & 20'' &53 & 20'' & 20'' & 40'' & 30'' & 20'' & 20'' & 25'' & 25''\\
24 & 25'' & 35'' & 25'' & 30'' & 40'' & 20'' & 15'' & 15'' &54 & 35'' & 20'' & 25'' & 40'' & 45'' & 30'' & 45'' & 65''\\
25 & - & 20'' & 10'' & 75'' & - & 25'' & 10'' & 40'' &55 & 30'' & 35'' & 25'' & 35'' & 25'' & 40'' & 30'' & 40''\\
26 & 15'' & 25'' & 15'' & 25'' & 15'' & 25'' & 20'' & 25'' &56 & 20'' & 15'' & 25'' & 10'' & 25'' & 25'' & 25'' & 30''\\
27 & 15'' & 20'' & 55'' & 65'' & 30'' & 20'' & 55'' & 25'' &57 & 50'' & 15'' & 20'' & 30'' & 35'' & 25'' & 45'' & 45''\\
28 & 20'' & 25'' & 20'' & 25'' & 20'' & 25'' & 35'' & 25'' &58 & 45'' & 40'' & 25'' & 50'' & 35'' & 30'' & 25'' & 30''\\
29 & 35'' & 40'' & 50'' & 45'' & 35'' & 35'' & 40'' & 45'' &59 & 20'' & 20'' & 20'' & 50'' & 20'' & 10'' & 30'' & 25''\\
30 & 30'' & 20'' & 20'' & 25'' & 25'' & 10'' & 35'' & 30'' &60 & 20'' & 20'' & 30'' & 40'' & 20'' & 20'' & 15'' & 15''\\
\enddata
\end{deluxetable*}

\begin{figure*}
    \centering
    \includegraphics[width=1.0\textwidth]{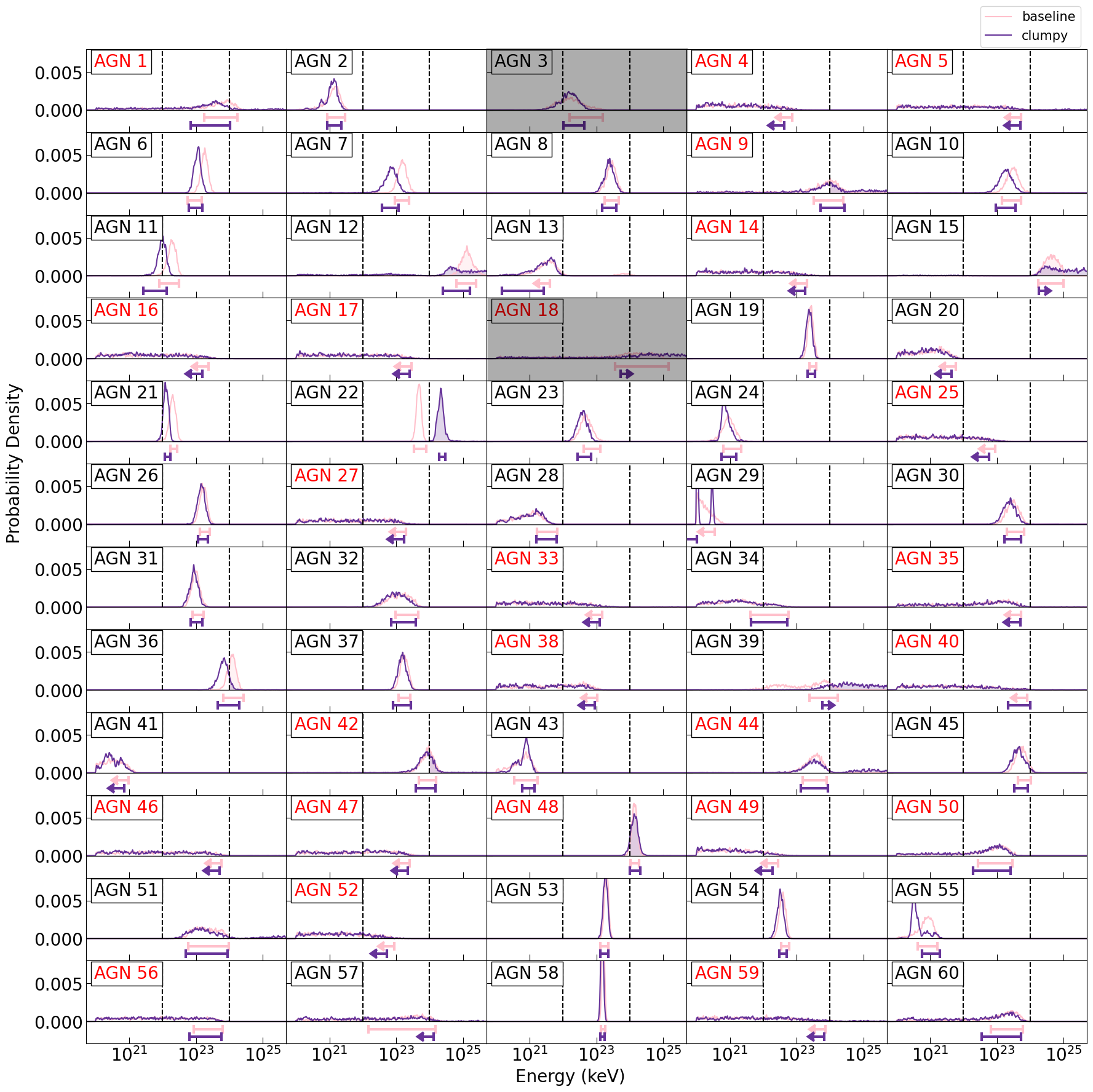}
    \caption{The posterior distributions for $N_{\mathrm{H}}$ found by BXA using both baseline (pink) and clumpy (purple) models. The vertical dashed line in each panel indicates the CT threshold, and CT portions of each posterior are shaded. The errorbar under the posterior distributions indicate the frequentest results for that source found using the standard \textsc{xspec} routines. Red labels indicate that XMM-Newton spectra did not exist for that source. Objects excluded from the analysis are shaded in gray.}
    \label{fig:bxa_all}
\end{figure*}

\mc{As the first step of spectral extraction, each source in the \citetalias{Zhao24-cycle5-6} catalog is assigned a circular region centered around its NuSTAR position.
The size of the region (radius $r$ in arc-seconds) is chosen to maximize the signal-to-noise ratio using the equation presented in \citet{Zappacosta18-NUSTAR_extragal_hard-band}:}

\begin{equation}
   \mc{ \text{SNR}(<r) = \frac{N(<r)}{\sqrt{N(<r) + 2B(<r)}}}
\end{equation}

\mc{Where $N(<r)$ is the total (not background-subtracted) number of counts extracted from a region with radius $r$ centered around the source, and $B(<r)$ is the number of counts extracted from the background maps produced by \textsc{nuskybgd}.
The SNR-maximized region sizes were determined separately for both FPMA and FPMB in each cycle.
When necessary, region sizes were manually adjusted to prevent overlap and avoid contamination from bright sources.
Table \ref{tab:snr_size} shows the resulting region sizes for all sources and observations.
A dash indicates that the source fell outside of  the FOV for every observation within that cycle.
All sources have extraction region sizes between $10 \arcsec$ and $75 \arcsec$, and $95 \%$ of source regions have radii $< 50 \arcsec$.}

\mc{We verified the spectral extraction process by comparing spectral counts to the photometric counts reported by \citetalias{Zhao24-cycle5-6}.
Because \citetalias{Zhao24-cycle5-6} uses $r = 20 ''$ for all sources in their analysis, spectral counts for this test were also extracted from $20 ''$ regions in every Cycle 5+6 observation.
Comparing net (background-subtracted) spectral counts to the net photometric counts yields a median difference (spec-phot) of $0.0^{+4.0}_{-2.0}$
and a median normalized difference (spec-phot/spec+phot) of $0.00^{+0.03}_{-0.02}$.
Therefore, we conclude that the net source counts in the spectra are consistent with the photometry.}

\section{Fit Details}\label{apendix-fits}

\begin{figure*}
    \centering
    \includegraphics[width=0.9\textwidth]{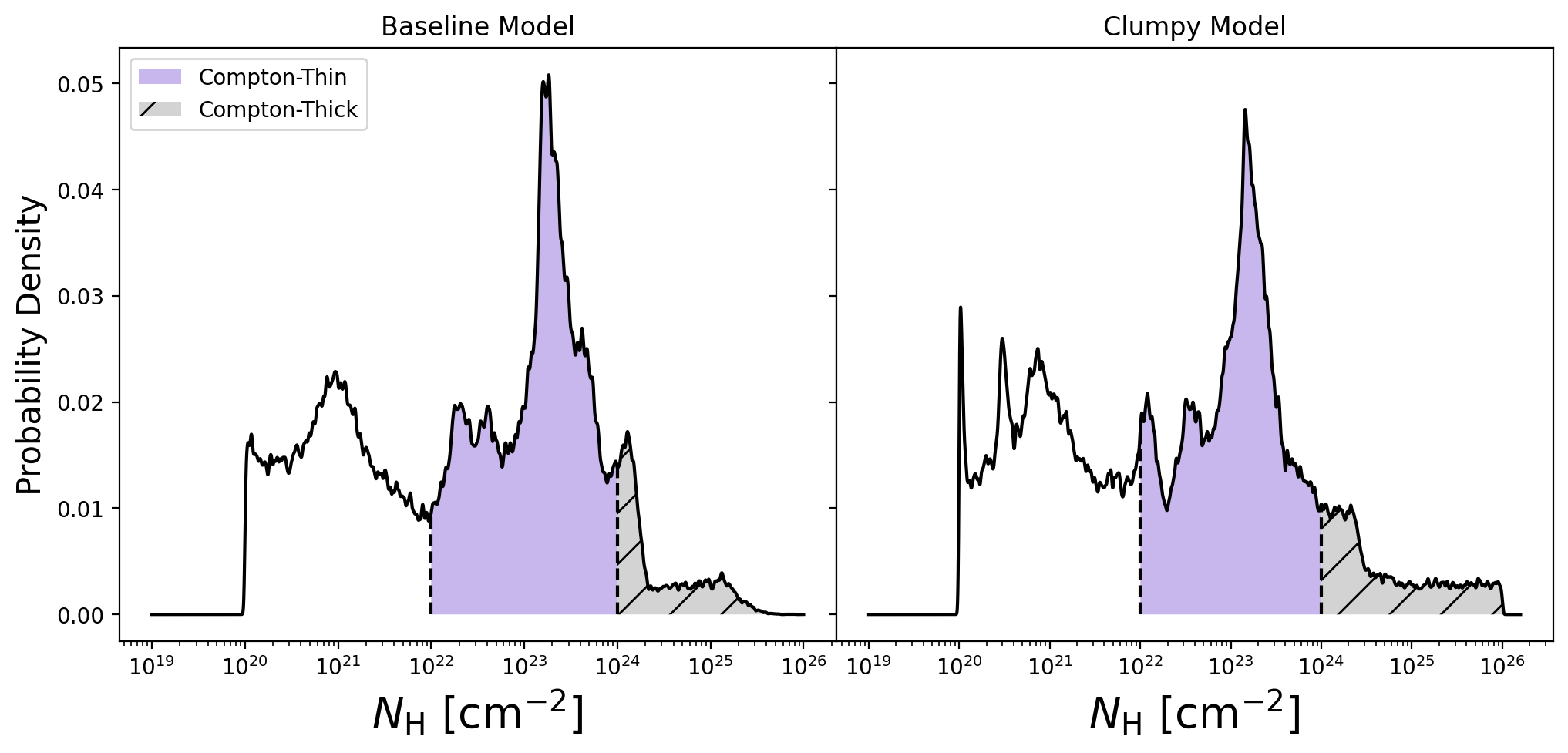}
    \caption{\mc{The added probability density functions (PDFs) of the sample for the baseline (left) and clumpy (right) models, giving the observed $N_{\mathrm{H}}$ distribution. The sum under the curve in the C-Thin (purple; solid) and CT (gray; hatched) regimes gives the effective C-Thin and CT sample size.}}
    \label{fig:nh_pdf}
\end{figure*}

The normalized PDFs derived from the BXA posteriors (see Section \ref{sec:BXA}) are shown in Figures \ref{fig:bxa_all} (individual PDFs) and \ref{fig:nh_pdf} (summed PDFs from the entire sample).
Table \ref{tab:pars} shows the best-fit parameters and $3 \sigma$ errors for each source.
Frozen parameters are given without error-bars.
Figure \ref{fig:models_all} shows the best-fit model (using \textsc{xspec}'s \textsc{eemodel}) and corresponding delta-chi residuals for each source.

\subsection{Notes on Individual Sources}

For both AGNs 36 and 54, there is an excess of emission in the FPMB spectrum at $<5$~keV that is not fit by the model. This excess does not appear in FPMA, and ignoring the excess bins has no significant effects on the fit.
Both objects have XMM data that are consistent with the best fit model, and $N_{\mathrm{H}}$ is well constrained (Figure \ref{fig:bxa_all}), so we conclude that the excess is not cause for concern.
 
When fitting AGN 60, we find that the XMM-Newton and NuSTAR data are inconsistent. 
The best-fit model matches the XMM-Newton data well, but it underestimates the NuSTAR spectra.
Similarly, when we fit exclusively to NuSTAR, the XMM-Newton spectra are overestimated by a factor of $\sim 3$. 
We conclude that the match between NuSTAR and XMM-Newton may be erroneous, so we only fit to the NuSTAR data.

\mc{Similarly, attempting to fit AGN 57 to both XMM-Newton and NuSTAR yields unphysical results. 
The best-fit models are CT with a lower limit on $N_{\mathrm{H}}$ (clumpy model) and an unconstrained scattered component (both models).
When XMM-Newton is excluded, a scattered component is not required and the constraints on $N_{\mathrm{H}}$ are reasonable, with upper error bars in the CT regime.
We conclude that the fit is more trustworthy when XMM-Newton is excluded.}

\subsubsection{CT Candidates}

Similar to AGNs 36 and 54, several of our CT candidates (AGNs 9, 18, and 48) had excess counts in the softest bins of one---but not both---of the NuSTAR instruments, and the excess counts do not contribute to the best-fit model.  
We discuss those objects in more detail here.

\textbf{AGN 9:} Visually, this object does not appear to be CT due to high counts at ~3 keV---the softest part of the NuSTAR spectrum--- for FPMA. 
However, the best fit model strongly prefers to be CT.
The CT classification appears to be driven by low counts (in both NuSTAR instruments) at $\sim$5 keV.
Since this object was not matched to any XMM sources, we conclude that more data is needed to accurately fit it, but we chose to include it in the analysis.

\textbf{AGN 18}: Similar to AGN 9, this object is faint and dominated by the background.
It does not visually appear to be CT, but the best-fit model strongly prefers high values of $N_{\mathrm{H}}$. 
However, $N_{\mathrm{H}}$ is poorly constrained for both the baseline and clumpy models (see Figure \ref{fig:bxa_all}) and fits for the galactic component are not consistent between the two (Figure \ref{fig:models_all}). 
For these reasons, we have decided that this fit is untrustworthy, and we do not include it in the analysis.

\textbf{AGN 48:} The excess emission for AGN 48 occurs in FPMA at $< 4.0$~keV. 
This emission does not affect the best fit, which is well constrained to be CT.
Further, when we ignore the bins with the excess counts, the spectral cutoff at $< 9 $~keV is visually clear, is consistent between FPMA and FPMB, and fits well to the model. 
Therefore we determine that this fit is trustworthy.


\startlongtable
\tabletypesize{\tiny}
\begin{deluxetable*}{lc|cccccc|cccccc}
\tablecaption{Best-Fit Model Parameters\label{tab:pars}}
\tiny
\startdata
\\
 &
 \multicolumn{6}{c}{\textbf{baseline}} &
  \multicolumn{6}{c}{\textbf{clumpy}} \\
  \hline
  & pass & & nH & & & & & & nH & & & &\\
  ID\tablenotemark{\tiny a} & (y/n)\tablenotemark{\tiny b}  & $\Gamma$ & ($10^{22}$~cm$^{-2}$) & $-$log($F$)\tablenotemark{\tiny c} & $n$\tablenotemark{\tiny d} & $f$\tablenotemark{\tiny e} & C/DOF & $\Gamma$ & ($10^{22}$~cm$^-2$) & $\kappa$\tablenotemark{\tiny f} & $n$\tablenotemark{\tiny d} & $f$\tablenotemark{\tiny e} & C/DOF \\
  \hline
1 & n & $1.80$ & $80.7_{-63}^{+93}$ & $13.3_{-0.4}^{+0.3}$ &  -  &  -  & 144.2/155 & $1.80$ & $39.9_{-33}^{+64}$ & $2.3_{-1.4}^{+3.9}$ &  -  &  -  & 144.7/155 \\
2 & n & $1.99_{-0.12}^{+0.13}$ & $0.2_{-0.1}^{+0.1}$ & $13.3_{-0.1}^{+0.1}$ & $1.1_{-0.5}^{+0.5}$ &  -  & 751.8/740 & $1.98_{-0.05}^{+0.13}$ & $0.2_{-0.1}^{+0.1}$ & $8.3_{-5}^{+0.7}$ & $8.3_{--1.4}^{+-5.8}$ &  -  & 743.8/740 \\
3 & n & $1.80$ & $4.4_{-2.9}^{+11}$ & $13.8_{-0.2}^{+0.2}$ &  -  & $\geq 2.6$ & 357.4/413 & $1.80$ & $2.1_{-1.1}^{+2.2}$ & $0.6_{-0.3}^{+0.2}$ & $0.6_{--0.3}^{+-0.5}$ &  -  & 355.1/413 \\
4 & n & $1.80$ & $\leq 7.3$ & $13.5_{-0.1}^{+0.1}$ &  -  &  -  & 622.5/646 & $1.80$ & $\leq 4.4$ & $1.8_{-0.4}^{+1.7}$ &  -  &  -  & 621.4/646 \\
\\
5 & n & $1.80$ & $\leq 53.4$ & $13.8_{-0.3}^{+0.4}$ &  -  &  -  & 472.7/518 & $1.80$ & $\leq 50.8$ & $0.5_{-0.2}^{+0.2}$ &  -  &  -  & 472.4/518 \\
6 & y & $1.48_{-0.26}^{+0.28}$ & $9.7_{-4.1}^{+5.5}$ & $13.4_{-0.1}^{+0.1}$ & $0.1_{-0.1}^{+0.1}$ &  -  & 598.4/600 & $1.6_{-0.22}^{+0.32}$ & $9.2_{-3.1}^{+6.5}$ & $2.5_{-1.1}^{+3.1}$ & $2.5_{--1.2}^{+-2.2}$ &  -  & 599.9/600 \\
7 & n & $1.80$ & $14.4_{-5.4}^{+9.4}$ & $13.3_{-0.1}^{+0.1}$ & $0.2_{-0.1}^{+0.1}$ &  -  & 477.0/559 & $1.80$ & $7_{-3.2}^{+5}$ & $1.5_{-0.5}^{+0.4}$ & $1.5_{--0.8}^{+-1.2}$ &  -  & 477.3/559 \\
8 & n & $1.80$ & $28.9_{-11}^{+17}$ & $13.5_{-0.2}^{+0.1}$ &  -  &  -  & 235.0/239 & $1.80$ & $25.5_{-10}^{+15}$ & $3.4_{-1.2}^{+1.4}$ &  -  &  -  & 235.7/239 \\
9 & n & $1.80$ & $109_{-75}^{+1.4e+02}$ & $13_{-0.3}^{+0.3}$ &  -  &  -  & 130.4/130 & $1.80$ & $93.6_{-42}^{+1.8e+02}$ & $10.1_{-6.2}^{+27}$ &  -  &  -  & 132.3/133 \\
10 & n & $1.80$ & $27.6_{-14}^{+26}$ & $13.8_{-0.2}^{+0.2}$ &  -  & $\geq 3.9$ & 476.8/526 & $1.80$ & $19.8_{-10}^{+17}$ & $1.5_{-0.6}^{+0.7}$ & $1.5_{--0.7}^{+-1.3}$ &  -  & 479.5/526 \\
11 & n & $1.72_{-0.43}^{+0.49}$ & $1.8_{-1}^{+1.3}$ & $13.9_{-0.2}^{+0.2}$ &  -  &  -  & 487.5/491 & $1.5_{-0.2}^{+0.31}$ & $0.8_{-0.5}^{+0.5}$ & $0.5_{-0.3}^{+0.2}$ &  -  &  -  & 472.6/491 \\
12 & n & $1.80$ & $1.2e+03_{-5.8e+02}^{+1.2e+03}$ & $13.7_{-0.1}^{+0.2}$ &  -  & $\geq 6.0$ & 568.7/598 & $1.80$ & $439_{-1.9e+02}^{+1.2e+03}$ & $40.4_{-12}^{+13}$ &  -  & $\geq 4.8$ & 584.6/598 \\
13 & n & $1.80$ & $\leq 0.4$ & $14_{-0.1}^{+0.1}$ &  -  & $\geq 0.0$ & 591.2/634 & $1.80$ & $0.1_{-0.1}^{+0.2}$ & $0.1_{-0.1}^{+0.1}$ &  -  &  -  & 589.5/635 \\
14 & n & $1.80$ & $\leq 20.3$ & $13.7_{-0.2}^{+0.2}$ &  -  &  -  & 469.3/524 & $1.80$ & $\leq 18.0$ & $0.7_{-0.2}^{+0.5}$ &  -  &  -  & 468.9/524 \\
15 & n & $1.80$ & $409_{-2.4e+02}^{+5.7e+02}$ & $13.5_{-0.2}^{+0.2}$ &  -  & $\geq 3.9$ & 250.4/276 & $1.80$ & $\geq 179.2$ & $19.7_{-6.3}^{+9.9}$ &  -  & $\geq 5.9$ & 257.3/276 \\
16 & n & $1.80$ & $\leq 23.4$ & $13.7_{-0.2}^{+0.2}$ &  -  &  -  & 395.2/430 & $1.80$ & $\leq 15.3$ & $0.4_{-0.1}^{+2.1}$ &  -  &  -  & 394.4/430 \\
17 & y & $1.80$ & $\leq 28.3$ & $13.5_{-0.2}^{+0.2}$ &  -  &  -  & 307.1/324 & $1.80$ & $\leq 25.4$ & $1.1_{-0.4}^{+1}$ &  -  &  -  & 306.8/324 \\
18 & n & $1.80$ & $123_{-87}^{+1.3e+03}$ & $13.1_{-0.5}^{+1.1}$ & $85.5_{-57}^{+63}$ &  -  & 357.1/353 & $1.80$ & $\geq 52.4$ & $15.5_{-14}^{+10}$ &  -  &  -  & 361.2/354 \\
19 & n & $1.80$ & $31.1_{-6.6}^{+8.3}$ & $13.5_{-0.1}^{+0.1}$ & $0.3_{-0.1}^{+0.2}$ &  -  & 884.3/914 & $1.80$ & $26.8_{-5.8}^{+8.4}$ & $6.1_{-0.9}^{+1.2}$ & $6.1_{--4.8}^{+-5.5}$ &  -  & 882.4/914 \\
20 & n & $1.7_{-0.28}^{+0.41}$ & $\leq 0.6$ & $14_{-0.2}^{+0.1}$ &  -  &  -  & 448.6/458 & $1.69_{-0.21}^{+0.3}$ & $\leq 0.4$ & $0.7_{-0.1}^{+1.3}$ &  -  &  -  & 446.5/458 \\
21 & y & $1.80$ & $2.2_{-0.5}^{+0.5}$ & $13.5_{-0}^{+0}$ &  -  & $\geq 6.8$ & 674.6/692 & $1.80$ & $1.5_{-0.3}^{+0.3}$ & $2.2_{-0.4}^{+0.5}$ & $2.2_{--1.6}^{+-2}$ &  -  & 677.7/692 \\
22 & y & $1.95_{-0.4}^{+0.49}$ & $52.7_{-18}^{+26}$ & $13.4_{-0.1}^{+0.2}$ &  -  &  -  & 461.8/456 & $1.80$ & $224_{-26}^{+70}$ & $56.5_{-8.7}^{+14}$ &  -  &  -  & 491.1/457 \\
23 & n & $1.80$ & $7.3_{-3.1}^{+5.8}$ & $13.7_{-0.1}^{+0.1}$ &  -  & $\geq 1.6$ & 427.9/488 & $1.80$ & $4.4_{-1.6}^{+2.6}$ & $1.2_{-0.3}^{+0.5}$ & $1.2_{--0.9}^{+-1.1}$ &  -  & 429.2/488 \\
24 & y & $1.72_{-0.09}^{+0.09}$ & $0.1_{-0.1}^{+0.1}$ & $13.3_{-0}^{+0}$ & $0.3_{-0.2}^{+0.3}$ &  -  & 714.5/723 & $1.73_{-0.08}^{+0.07}$ & $0.1_{-0}^{+0.1}$ & $2.7_{-0.9}^{+2.4}$ & $2.7_{--1.4}^{+-2}$ &  -  & 714.2/723 \\
25 & y & $1.80$ & $\leq 9.0$ & $13.4_{-0.1}^{+0.1}$ &  -  &  -  & 394.5/379 & $1.80$ & $\leq 5.8$ & $0.8_{-0.2}^{+2.1}$ &  -  &  -  & 394.0/379 \\
26 & n & $1.80$ & $18.5_{-5.2}^{+7.3}$ & $13.7_{-0.1}^{+0.1}$ &  -  &  -  & 461.7/456 & $1.80$ & $16.6_{-5}^{+6.4}$ & $4.6_{-1}^{+1.2}$ &  -  &  -  & 459.4/456 \\
27 & n & $1.80$ & $\leq 19.8$ & $13.6_{-0.2}^{+0.2}$ &  -  &  -  & 659.2/740 & $1.80$ & $\leq 17.1$ & $0.6_{-0.2}^{+0.5}$ &  -  &  -  & 658.6/740 \\
28 & y & $1.80$ & $0.4_{-0.2}^{+0.3}$ & $13.8_{-0.1}^{+0.1}$ & $0.3_{-0.1}^{+0.1}$ &  -  & 850.0/826 & $1.80$ & $0.3_{-0.2}^{+0.3}$ & $1.1_{-0.2}^{+0.6}$ & $1.1_{--0.4}^{+-0.4}$ &  -  & 845.8/826 \\
29 & y & $1.62_{-0.02}^{+0.02}$ & $\leq 0.0$ & $12.7_{-0}^{+0}$ & $1.5_{-0.7}^{+1}$ &  -  & 2060.5/2050 & $1.69_{-0.01}^{+0.02}$ & $\leq 0.0$ & $22_{-0.8}^{+2.3}$ &  -  & $\geq 0.0$ & 2047.8/2050 \\
30 & n & $1.80$ & $38.4_{-18}^{+27}$ & $13.7_{-0.2}^{+0.1}$ &  -  &  -  & 515.3/520 & $1.80$ & $32_{-15}^{+22}$ & $4.6_{-1.6}^{+2.2}$ &  -  &  -  & 514.4/520 \\
31 & y & $1.80$ & $11.4_{-3.5}^{+5.9}$ & $13.4_{-0.1}^{+0.1}$ &  -  &  -  & 485.8/441 & $1.80$ & $10.5_{-3.6}^{+4.9}$ & $2.5_{-0.4}^{+0.5}$ &  -  &  -  & 483.9/441 \\
32 & n & $1.80$ & $26.6_{-17}^{+20}$ & $14_{-0.2}^{+0.1}$ &  -  &  -  & 658.8/760 & $1.80$ & $22.8_{-16}^{+15}$ & $2.6_{-1.1}^{+1.1}$ &  -  &  -  & 661.8/765 \\
33 & n & $1.80$ & $\leq 15.0$ & $13.7_{-0.2}^{+0.2}$ &  -  &  -  & 214.5/203 & $1.80$ & $\leq 12.8$ & $1.1_{-0.6}^{+0.8}$ &  -  &  -  & 214.3/203 \\
34 & n & $1.80$ & $2.2_{-1.8}^{+3.5}$ & $14.2_{-0.3}^{+0.2}$ & $0.1_{-0}^{+0}$ &  -  & 401.3/454 & $1.80$ & $2.6_{-2.1}^{+2.8}$ & $0.5_{-0.3}^{+0.2}$ & $0.5_{--0.1}^{+-0.3}$ &  -  & 400.3/454 \\
35 & n & $1.80$ & $\leq 53.5$ & $13.5_{-0.3}^{+0.3}$ &  -  &  -  & 259.6/247 & $1.80$ & $\leq 50.7$ & $1.1_{-0.6}^{+1.5}$ &  -  &  -  & 259.6/247 \\
36 & n & $2.06_{-0.82}^{+1.1}$ & $152_{-87}^{+1.2e+02}$ & $13.5_{-0.3}^{+0.3}$ &  -  & $\geq 0.1$ & 738.4/733 & $2.17_{-0.54}^{+0.81}$ & $94.3_{-50}^{+1e+02}$ & $\leq 271.3$ &  -  & $\geq 0.1$ & 745.5/735 \\
\\
37 & n & $1.80$ & $17.6_{-5.7}^{+8.8}$ & $13.6_{-0.1}^{+0.1}$ &  -  &  -  & 537.2/575 & $1.8_{-0.42}^{+0.58}$ & $15.4_{-7.4}^{+13}$ & $\leq 13.7$ &  -  &  -  & 537.7/574 \\
38 & n & $1.80$ & $\leq 10.8$ & $12.7_{-0.1}^{+0.1}$ &  -  &  -  & 297.3/299 & $1.80$ & $\leq 9.0$ & $\leq 8.3$ &  -  &  -  & 296.1/299 \\
39 & y & $1.80$ & $82.8_{-58}^{+92}$ & $13.3_{-0.3}^{+0.2}$ &  -  & $\geq 2.0$ & 242.0/238 & $1.80$ & $\geq 58.5$ & $\leq 90.2$ &  -  &  -  & 250.6/239 \\
40 & y & $1.80$ & $\leq 81.3$ & $13.3_{-0.1}^{+0.6}$ &  -  &  -  & 187.7/229 & $1.80$ & $51.1_{-29}^{+48}$ & $2.9_{-1.4}^{+2.7}$ &  -  & $\geq 3.2$ & 187.5/229 \\
41 & y & $1.56_{-0.12}^{+0.19}$ & $\leq 0.1$ & $13.9_{-0.1}^{+0.1}$ &  -  &  -  & 774.9/794 & $1.58_{-0.1}^{+0.09}$ & $\leq 0.1$ & $0.9_{-0.1}^{+0.7}$ &  -  &  -  & 772.0/794 \\
42 & y & $1.80$ & $94.4_{-46}^{+65}$ & $13.1_{-0.3}^{+0.2}$ & $\leq 41.8$ &  -  & 339.2/376 & $1.80$ & $79.4_{-41}^{+71}$ & $3.4_{-2}^{+4}$ & $3.4_{--1.4}^{+52}$ &  -  & 340.2/376 \\
43 & y & $2.08_{-0.1}^{+0.11}$ & $0.1_{-0.1}^{+0.1}$ & $13.7_{-0.1}^{+0}$ &  -  &  -  & 1197.5/1150 & $2.08_{-0.07}^{+0.08}$ & $0.1_{-0}^{+0.1}$ & $5_{-2}^{+2.9}$ &  -  &  -  & 1186.3/1150 \\
44 & y & $1.80$ & $38.9_{-24}^{+41}$ & $13.5_{-0.2}^{+0.2}$ &  -  &  -  & 453.8/503 & $1.80$ & $36.2_{-23}^{+49}$ & $1.5_{-0.7}^{+1.8}$ &  -  &  -  & 453.0/503 \\
45 & y & $1.80$ & $68.3_{-26}^{+36}$ & $13.4_{-0.1}^{+0.1}$ &  -  &  -  & 546.2/543 & $1.80$ & $49.7_{-16}^{+36}$ & $7.1_{-1.8}^{+1.8}$ &  -  &  -  & 544.9/543 \\
46 & y & $1.80$ & $\leq 58.7$ & $13.7_{-0.2}^{+0.2}$ &  -  &  -  & 297.6/337 & $1.80$ & $\leq 50.9$ & $1.5_{-0.7}^{+1.3}$ &  -  &  -  & 296.7/337 \\
47 & y & $1.80$ & $\leq 25.7$ & $13.7_{-0.2}^{+0.2}$ &  -  &  -  & 520.8/603 & $1.80$ & $\leq 22.7$ & $0.5_{-0.1}^{+0.7}$ &  -  &  -  & 520.3/603 \\
48 & y & $1.80$ & $141_{-36}^{+46}$ & $12.4_{-0.1}^{+0.1}$ &  -  &  -  & 290.2/300 & $1.80$ & $144_{-46}^{+61}$ & $24.2_{-9.9}^{+14}$ &  -  &  -  & 292.0/300 \\
49 & n & $1.80$ & $\leq 2.8$ & $13.4_{-0.1}^{+0.1}$ &  -  &  -  & 249.3/278 & $1.80$ & $\leq 1.9$ & $3.2_{-2.6}^{+0.9}$ &  -  &  -  & 248.6/278 \\
50 & n & $1.80$ & $13.1_{-10}^{+16}$ & $13.5_{-0.2}^{+0.2}$ &  -  &  -  & 363.0/390 & $1.80$ & $10.9_{-9}^{+15}$ & $1.2_{-0.4}^{+0.7}$ &  -  &  -  & 363.5/390 \\
51 & n & $1.80$ & $15.6_{-9.7}^{+80}$ & $13.4_{-0.2}^{+0.3}$ &  -  &  -  & 289.7/318 & $1.80$ & $12.3_{-7.4}^{+75}$ & $2.5_{-0.7}^{+3.3}$ &  -  &  -  & 288.3/318 \\
52 & n & $1.92_{-0.73}^{+0.97}$ & $\leq 8.7$ & $13.7_{-0.2}^{+0.3}$ &  -  &  -  & 335.1/358 & $1.80$ & $\leq 5.4$ & $0.6_{-0.2}^{+1.2}$ &  -  &  -  & 335.4/359 \\
53 & y & $1.46_{-0.19}^{+0.2}$ & $17.2_{-4.2}^{+5.4}$ & $13.4_{-0.1}^{+0.1}$ &  -  &  -  & 784.4/771 & $1.6_{-0.18}^{+0.2}$ & $17.2_{-4.3}^{+5.6}$ & $3.2_{-1.3}^{+2.6}$ &  -  &  -  & 784.4/771 \\
54 & y & $1.80$ & $4.5_{-1.1}^{+1.3}$ & $13.8_{-0.1}^{+0.1}$ & $0.1_{-0}^{+0.1}$ &  -  & 926.2/993 & $1.80$ & $4_{-1}^{+1}$ & $1.7_{-0.3}^{+0.3}$ & $1.7_{--1.3}^{+-1.5}$ &  -  & 924.9/993 \\
55 & y & $1.66_{-0.05}^{+0.05}$ & $0.1_{-0.1}^{+0.1}$ & $13_{-0}^{+0}$ & $0.8_{-0.4}^{+0.4}$ &  -  & 1277.7/1318 & $1.72_{-0.04}^{+0.03}$ & $0.2_{-0.1}^{+0}$ & $7.6_{-3.7}^{+1.1}$ & $7.6_{--2.1}^{+-5.4}$ &  -  & 1265.7/1318 \\
56 & n & $1.80$ & $33.4_{-25}^{+30}$ & $13.4_{-0.2}^{+0.2}$ & $57.1_{-37}^{+37}$ &  -  & 353.2/359 & $1.80$ & $28.6_{-22}^{+28}$ & $1.4_{-0.7}^{+1.1}$ & $1.4_{-76}^{+1.2e+02}$ &  -  & 353.6/359 \\
57 & y & $1.80$ & $54.3_{-53}^{+98}$ & $13.4_{-0.2}^{+0.1}$ &  -  &  -  & 456.0/552 & $1.80$ & $\leq 132.8$ & $7.4_{-4}^{+6.4}$ &  -  &  -  & 455.5/552 \\
58 & y & $1.83_{-0.14}^{+0.15}$ & $16.2_{-2.3}^{+2.6}$ & $12.7_{-0}^{+0.1}$ &  -  &  -  & 1026.6/1095 & $1.92_{-0.15}^{+0.15}$ & $15.2_{-2.3}^{+2.7}$ & $12.1_{-3.7}^{+6}$ &  -  &  -  & 1020.2/1095 \\
59 & n & $1.80$ & $\leq 74.0$ & $13.8_{-0.2}^{+0.5}$ &  -  &  -  & 360.7/346 & $1.80$ & $\leq 68.4$ & $\leq 1.1$ &  -  &  -  & 360.4/346 \\
60 & n & $1.80$ & $30.5_{-24}^{+30}$ & $13.3_{-0.2}^{+0.1}$ &  -  &  -  & 290.9/312 & $1.80$ & $26.1_{-23}^{+27}$ & $3.4_{-1.3}^{+1.8}$ &  -  &  -  & 290.7/312 \\
\enddata
\tablecomments{ \\
$^{a}$ NuSTAR IDs from the catalog by \citetalias{Zhao24-cycle5-6} \\
$^{b}$ Whether this object passed the 8-24 keV signal-to-noise and maximum likelihood requirements to be included in the Section \ref{sec:nh_int} analysis. \\
$^{c}$ $-\log(F_{2-10}/\mathrm{erg\,cm^{-2}\,s^{-1}})$: Flux from 2–10 keV. \\
$^{d}$ Normalization of the \textsc{mekal} model: $\frac{10^{-14}}{4 \pi [D_A (1 + z)]^{2}} \int n_e n_H dV$ \\
$^{e}$ Factor of scattered power law emission (omni component of \textsc{uxclumpy}) \\
$^{f}$ Normalization of the \textsc{powerlaw} model: $10^{-3}$~photons~keV$^{-1}$~cm$^{-2}$s$^{-1}$ at 1 keV
}
\end{deluxetable*}

\begin{figure*}[h!]
    \centering
    \includegraphics[width=1.0\textwidth]{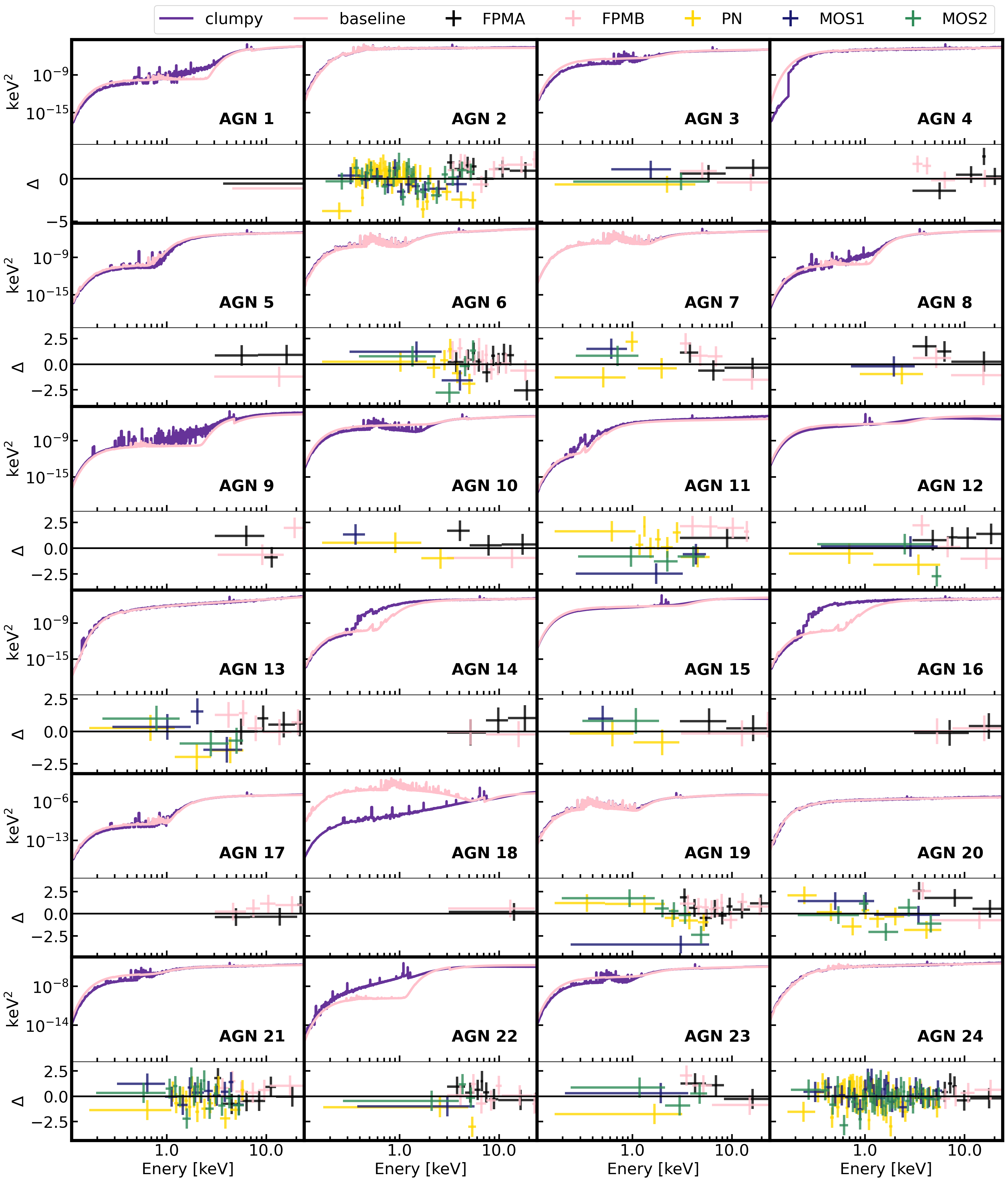}
\end{figure*}%
\begin{figure*}[ht]
    \includegraphics[width=1.0\textwidth]{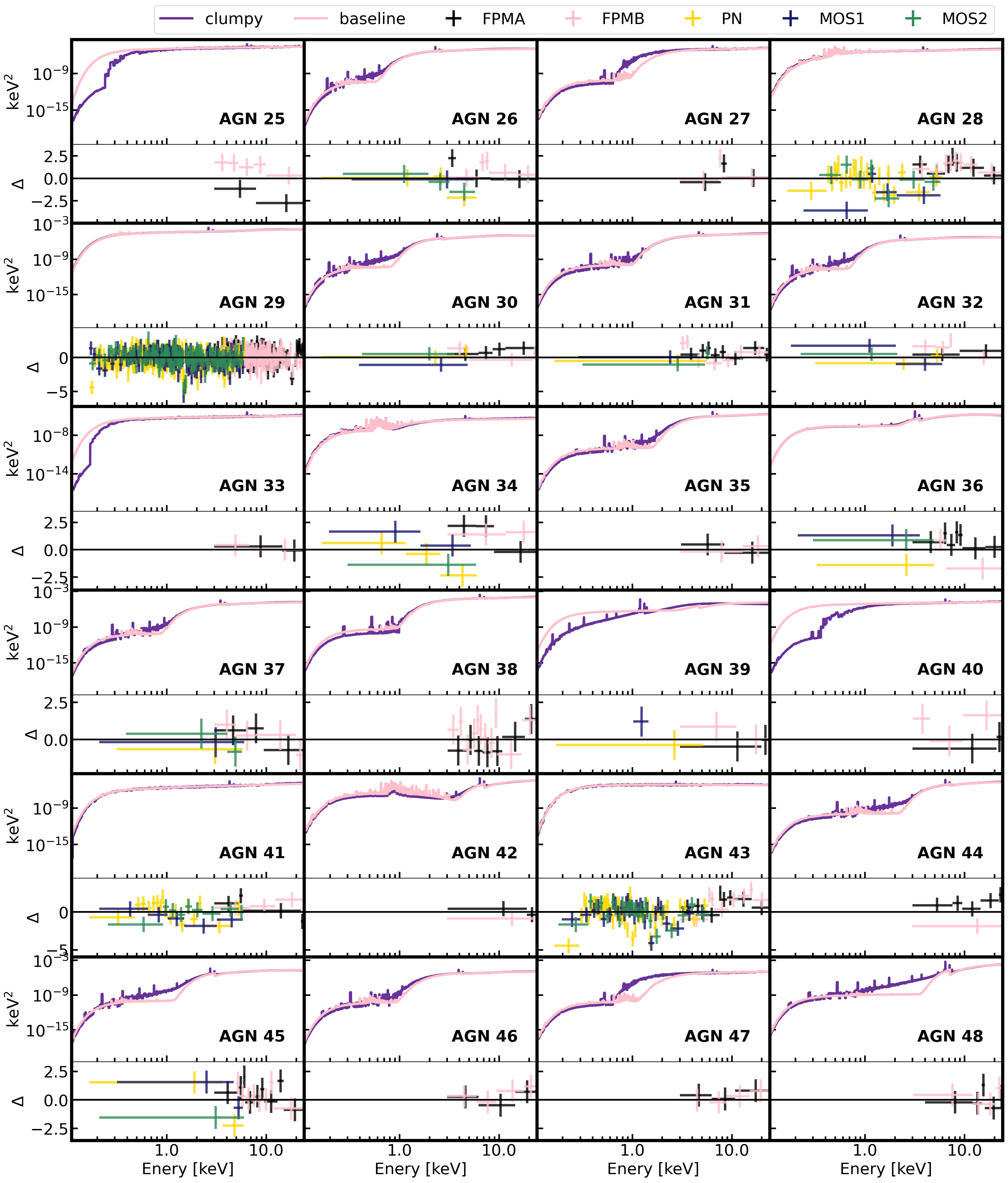}
\end{figure*}%
\begin{figure*}[ht]
    \includegraphics[width=1.0\textwidth]{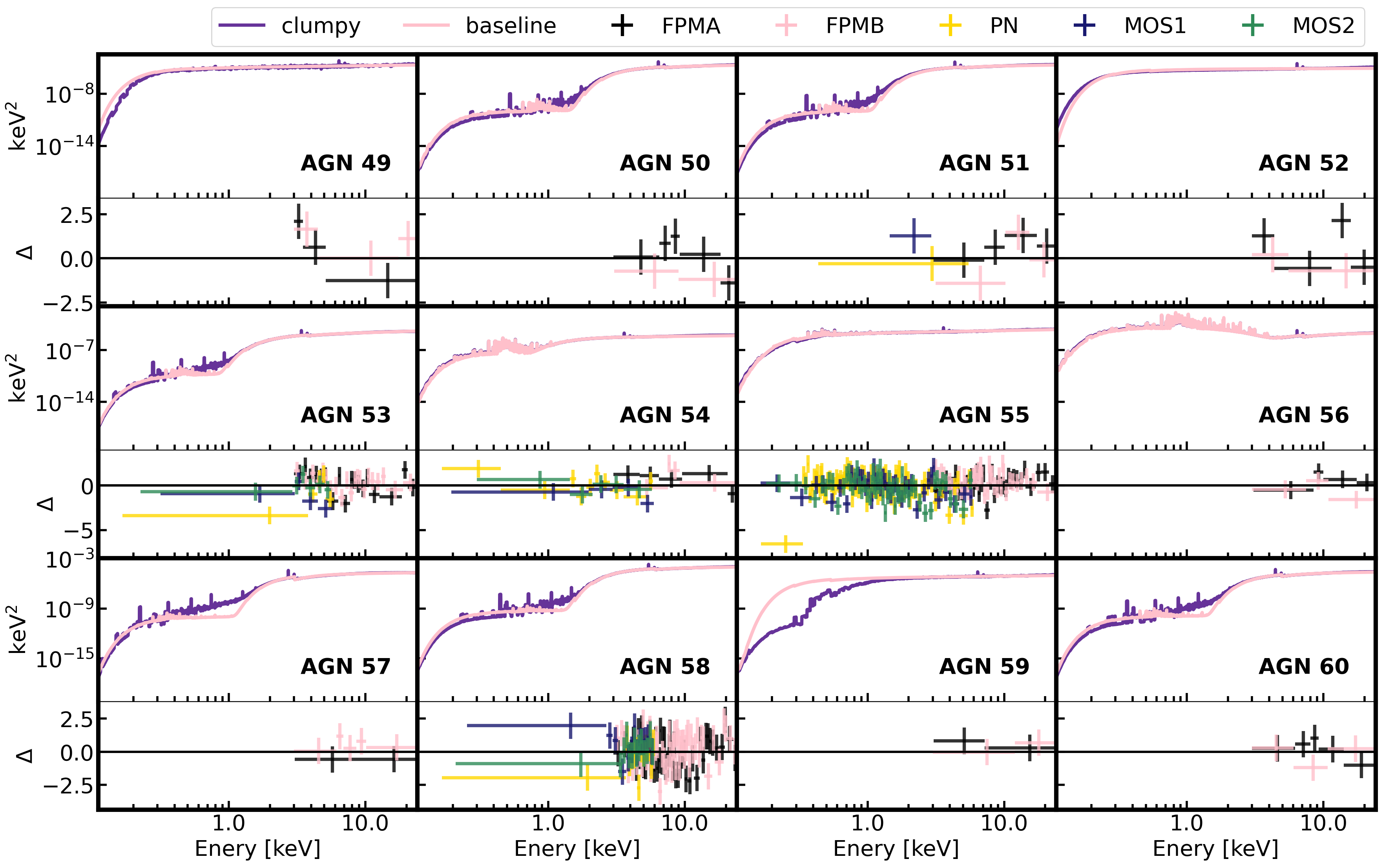}
    \caption{Best-fit clumpy (purple) and baseline (pink) \textsc{eemodel} for each source compared to the delta-chi error ($\Delta$).}
    \label{fig:models_all}
\end{figure*}
\end{document}